\documentclass[12pt]{iopart}

\pdfoutput=1

\expandafter\let\csname equation*\endcsname\relax
\expandafter\let\csname endequation*\endcsname\relax
\usepackage{amsmath}
\usepackage{amssymb}
\usepackage{graphicx}% Include figure files
\usepackage{dcolumn}% Align table columns on decimal point
\usepackage{bm}% bold math
\newcommand{\bra}[1]{\ensuremath{\left\langle #1\right|}}
\newcommand{\ket}[1]{\ensuremath{\left|#1\right\rangle}}

\newcommand{\mean}[1]{\ensuremath{\left\langle #1\right\rangle}}

\begin{document}
\graphicspath{{pictures/}}

\title{Dual-probe decoherence microscopy: \\ Probing pockets of coherence in a decohering environment}

\author{Jan Jeske$^{1,2,3}$, Jared H. Cole$^{1,2,3}$, Clemens M\"uller$^{4,5}$, Michael Marthaler$^{2,3}$, Gerd Sch\"on$^{2,3}$}
%\ead{jan.jeske@student.rmit.edu.au}
\address{$^1$Chemical and Quantum Physics, School of Applied Sciences, RMIT University, Melbourne 3001, Australia}
\address{$^2$Institut f\"ur Theoretische Festk\"{o}rperphysik, Karlsruhe Institute of Technology, 
D-76128 Karlsruhe, Germany}
\address{$^3$DFG-Center for Functional Nanostructures (CFN), D-76128 Karlsruhe, Germany}
\address{$^4$Institut f\"{u}r Theorie der Kondensierten Materie, Karlsruhe Institute of Technology, D-76128 Karlsruhe, Germany}
\address{$^5$D\'epartement de Physique, Universit\'e de Sherbrooke, Sherbrooke, Qu\'ebec, Canada J1K 2R1}

%\date{\today}

\begin{abstract}
We study the use of a pair of qubits as a decoherence probe of a non-trivial environment.  This dual-probe configuration is modelled by three two-level-systems which are coupled in a chain in which the middle system represents an \emph{environmental} two-level-system (TLS).  This TLS resides within the environment of the qubits and therefore its coupling to perturbing fluctuations (i.e.~its decoherence) is assumed much stronger than the decoherence acting on the probe qubits. We study the evolution of such a tripartite system including the appearance of a decoherence-free state (dark state) and non-Markovian behaviour. We find that all parameters of this TLS can be obtained from measurements of one of the probe qubits. Furthermore we show the advantages of two qubits in probing environments and the new dynamics imposed by a TLS which couples to two qubits at once.
\end{abstract}

\maketitle

\section{Introduction} 
The loss of coherence (decoherence) of quantum bits (qubits) due to environmental perturbations is an important obstacle on the way to large scale quantum electronics and quantum computation. Such perturbations at the same time contain information about the surrounding environment which generates them. The idea of using qubits as probes of their environment has recently generated interest~\cite{ShnirmanSchoen2008, Coledecoherencemicroscopy, Chernobrod2005, Weber2008, Sousa2009} as an alternative application of qubit technology where the effects of decoherence are used, rather than suppressed. 

In general, when an environment acts on a qubit as a weakly coupled, fluctuating bath, the environmental effects can be simply expressed as a relaxation and an excitation rate as well as a pure dephasing rate. The decoherence process becomes much more complex when a qubit couples to any component of an environment strongly enough such that quantum mechanical levels within the environment need to be taken into account.  In many systems, such partially coherent ``pockets'' are observed in the environment.  Examples include two-level fluctuators in superconducting devices~\cite{Martinis2005, Neeley2008, Martin2005, Bushev2010, Cole2010defectmodels, Lisenfeld2010}, impurity spins in semiconductors~\cite{Jelezko2004, Jelezko2004a, Gaebel2006, Childress2006, Morello2010} and single-molecule magnets\cite{Barco2000, Awschalom1992, Tejada1997} such as $^8$Fe and Ferritin.
Valuable information about the quantum mechanical nature of the environment may be obtained by using qubits as a direct environmental probe, with potentially high sensitivity and high spatial resolution~\cite{Coledecoherencemicroscopy}.

The concept of qubit probes has already been used to realise a nano-magnetometer using Nitrogen-Vacancy (NV) centres in diamond attached to the end of atomic force microscope cantilevers~\cite{Balasubramanian2008, Maze2008, Degen2008, Taylor2008, Balasubramanian2009}.  In this case the Zeeman splitting of electronic levels within the NV centre is optically probed to provide a direct measure of the local magnetic field to nanometre resolution~\cite{Balasubramanian2009}.  Additional information which one may obtain from the decoherence processes has also been studied\cite{Hall2009, Hall2010} in this context of classically fluctuating fields.

We investigate a more general configuration in which a decoherence probe couples transversally to a localized \emph{pocket of coherence} and can therefore exchange energy with, and extract information from, its immediate environment. We model such a pocket as an environmental two-level-system (TLS), for example a charge- or spin-impurity, which is then in turn coupled to its surrounding environment.  We will show a full solution of the system dynamics in the relevant parameter regimes and demonstrate how one could use this information to identify and characterize an environmental TLS.

The behaviour of one qubit interacting with such a partially coherent component of the environment is well understood~\cite{Mueller2009}. In such systems ambiguity often arises between oscillatory evolution induced by partially coherent defects and oscillatory evolution which stems from the single qubit Hamiltonian itself. Two non-interacting qubits which couple to the same environment (figure \ref{fig system}\textbf{A}) enable the observation of any environmentally induced correlations between these qubits, removing the ambiguity. This indirect interaction is a good indicator for the existence of coherent regions in the closer environment. 

Solving the equations of motion for such a model numerically results in a variety of complex behaviours~\cite{Oxtoby2009, Yuan2008, Emary2008, Paladino2008} in different regimes of parameter space. 
In this paper we focus on solving the system analytically in key regimes and then use these analytical solutions to understand the more general behaviour numerically. 

Our model of two qubits coupled to a common environmental TLS has many similarities to the well studied problem of two qubits coupled to a quantum harmonic oscillator.  This classic Tavis-Cummings model has been widely investigated for its use in quantum information processing and studying atom-photon interactions.  Interpreted in this context our results complement a variety of physical effects which appear in the Tavis-Cummings model, including dispersive coupling~\cite{Raimond2001, Sorensen1999, Zheng2000, Majer2007, Filipp2009, Niskanen2006}, ultra-strong coupling~\cite{Niemczyk2010, Fink2009, Casanova2010} and entanglement birth and death~\cite{Yu2009, Yonac2006, Yu2007, Cole2010entanglementsuddendeath}.  It is natural that similar effects appear in both systems due to their formal equivalence within the single-excitation subspace.  There are however important physical differences, as an environmental TLS is typically more localised in space than a quantum harmonic oscillator and the symmetry of the TLS allows a more general coupling to its environment.

After introducing the theoretical model in section \ref{model} we present a general analysis of the system in section \ref{sec general physics}: 
In the weak decoherence regime we find an oscillation of energy between the qubits, i.e. an environmentally mediated coupling. We find a decoherence-free state (dark state), which leads to formation of stray entanglement between the qubits in the decoherence process independent of the decoherence strength of the TLS. In our analytical solutions we find a clear threshold between oscillations and decay. In order to clarify what constitutes an environmental pocket of coherence it is shown that the same threshold divides Markovian from non-Markovian system dynamics.  We define an effective decay rate of the qubit dynamics which turns out to have a linear dependence on the decoherence rates of the TLS in the weak decoherence regime and a roughly inverse dependence in the strong decoherence regime.
In section \ref{sec probing} we interpret our results in the context of dual-probe microscopy and we find that in the weak decoherence regime an environmental TLS can be fully characterized and located in a substrate. In the strong decoherence regime the TLS can only be located. Section \ref{experimental realisations} puts the theoretical model in context of present experimental qubit realizations.

\section{Model and methods \label{model}}

\begin{figure*}
\includegraphics[scale=1]{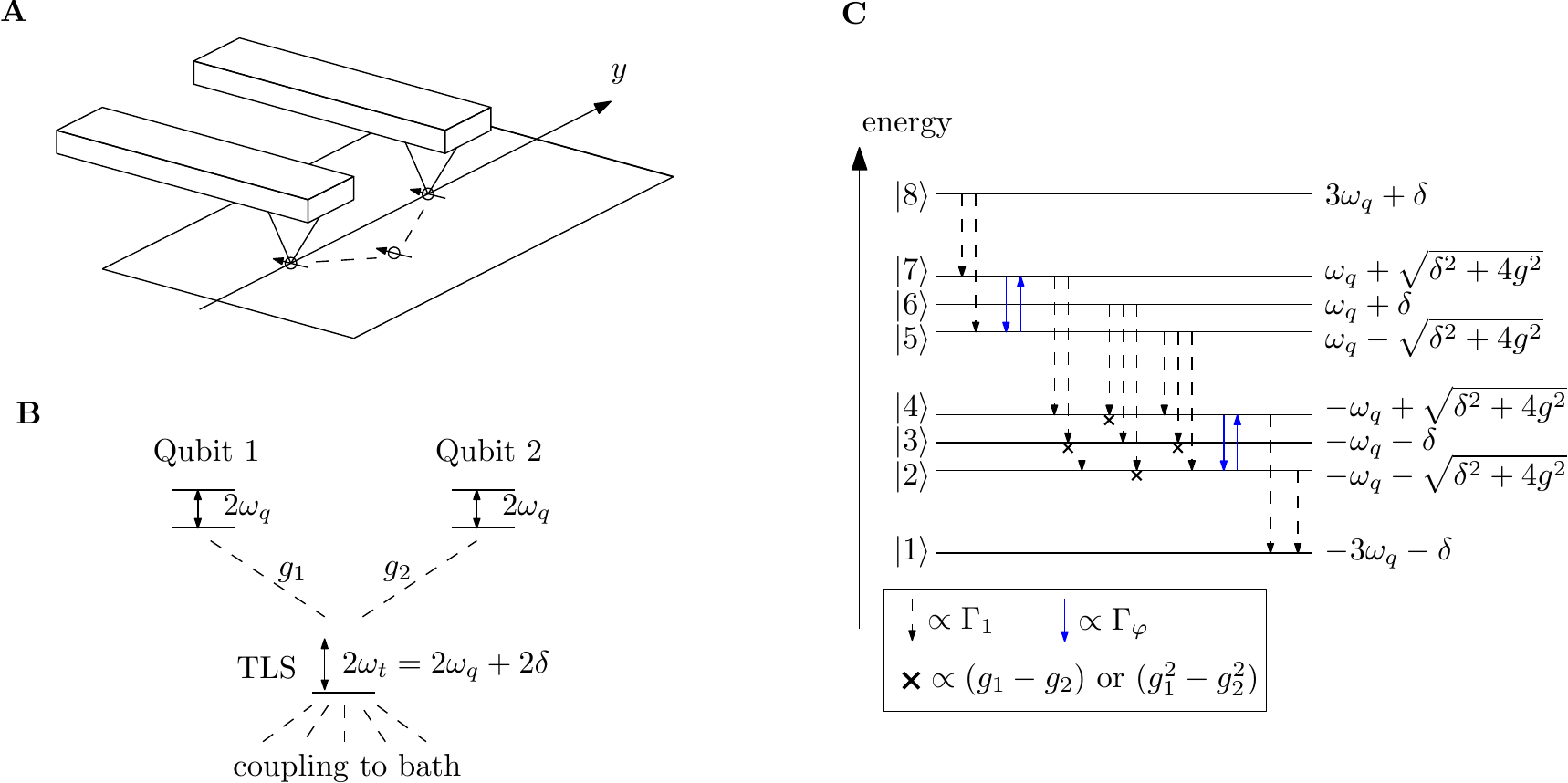}
\caption{\textbf{A}  Experimental setup in which two qubits are attached to an atomic force microscope and moved along their connecting line $y$ with a fixed distance between them. The TLS is located in the substrate underneath the qubits.  \; \textbf{B}  Illustration of the model system investigated in this paper: Two qubits with level splitting $\omega_q$ coupled with individual coupling strengths $g_1, g_2$ to an environmental TLS detuned by $2\delta$. The TLS in turn is coupled to a bath.\; \textbf{C} Energies of the Hamiltonian eigenstates and the corresponding decohering transitions. Dephasing transitions ($\propto \Gamma_\varphi$) transfer population between states of similar energies, relaxation transitions ($\propto \Gamma_1$) transfer population to subspaces of lower excitation number. Crosses indicate rates which disappear for $g_1=g_2$. The eigenstates are given in appendix \ref{app eigenstates} 
\label{fig system} }%rates from Notizblock 4 (Diplomzeit) and "full hilbert space.nb"
\end{figure*}

In order to study how a qubit probe pair interacts with an environmental TLS we construct a simplified model.  Consider an experimental setup in which the two qubits are attached to an atomic force microscope such that they can be positioned precisely (with a fixed distance between them) on top of a scanned substrate which contains the TLS (figure \ref{fig system}\textbf{A}). At several positions the population of the excited state of the qubit is measured as a function of time. In each position of the cantilever the coupling strengths between the qubits and the TLS, $g_1$ and $g_2$, vary due to their relative position. We will give a detailed model of this variation in Sec. \ref{sec probing}. The full system Hamiltonian can be written as:
\begin{align}
H_{sys} &=  \omega_q \sigma_z^{Q1} + \omega_q \sigma_z^{Q2} + (\omega_q + \delta )\sigma_z^{TLS} 
  + g_1 (\sigma_x^{Q1} \sigma_x^{TLS} + \sigma_y^{Q1} \sigma_y^{TLS}) \nonumber \\ 
  & \;\; + g_2 (\sigma_x^{Q2} \sigma_x^{TLS} + \sigma_y^{Q2} \sigma_y^{TLS}) \label{system Hamiltonian}
\end{align}
where $\sigma_x, \sigma_y$ and $\sigma_z$ are the respective Pauli operator which act on qubit 1 (Q1), qubit 2 (Q2) or the TLS (TLS). The first two terms describe the two qubits which have the same level splitting of $2\omega_q$ while the third term describes the TLS with a level splitting of $2(\omega_q+\delta)$. Here, $\delta$ is the relative detuning between qubits and TLS. The last two terms in eq. \eqref{system Hamiltonian} are transversal coupling terms between each of the qubits and the TLS with the respective coupling strengths $g_1$ and $g_2$ and we introduce $g=\sqrt{g_1^2+g_2^2}$ for later simplicity. We focus on the specific case of transversal coupling as we are particularly interested in direct energy exchange between the qubits and the TLS. Throughout this discussion, we use the terminology ``qubit'' and ``TLS'' to differentiate between the fabricated and controllable two-state-probes and the environmental two-level-system of interest. For ease of notation we will use the convention $\hbar = k_B  = 1$.

As the system Hamiltonian $H_{sys}$ is block-diagonal, the coherent evolution is limited to the subspace states with equal excitation number. For the time evolution, we choose the state in which qubit 1 is in its excited state and the other two subsystems are in their ground state $|Q1,Q2,TLS\rangle = |\uparrow \downarrow \downarrow \rangle$ as the system's initial state. This is a state with a single excitation, and therefore we can neglect the subspaces of higher excitation numbers in the following calculations. 

A key advantage of probe qubits attached to a cantilever is that they can be calibrated while lifted away from the sample.  This allows the \emph{intrinsic} decoherence of the probes themselves to be accurately characterized. Once the probes are brought close to the sample, this intrinsic decoherence defines a limit in the sensitivity for detecting features within the sample. For our purposes, an ideal qubit probe is one with a very long intrinsic decoherence time, as this provides a large dynamic range for sensing.  We assume the intrinsic contribution to decoherence to be small compared to the dynamics induced by the environmental TLS. Then we can ignore this contribution in what follows, i.e.~assume that only the TLS is coupled to a fluctuating bath. For its coupling to the environment we take the operator 
\begin{equation}
	H_{int}=\hat s \hat B=(v_\perp \sigma_x^{TLS} + v_\parallel \sigma_z^{TLS}) \hat B
	\label{eq:HInt}
\end{equation}
where $v_\perp$ and $v_\parallel$ are the transversal and longitudinal coupling strengths respectively and $\hat B$ is an operator acting on the bath. We assume a low temperature bath $\omega_q \gg T $. 

In most qubit architectures, the qubit's level splitting is significantly larger than the other energy scales in the problem.  Typically this is a requirement to obtain coherence over long time scales as well as adequate control over the quantum system.  We therefore assume throughout this discussion that $\omega_q \gg \delta, g_1, g_2, T$. Under this assumption we can neglect subspaces with more than one excitation. A large $\omega_q$ guarantees a clear separation of these subspaces in energy while a low temperature bath guarantees the absence of spontaneous excitations from the bath. In the limit of large $\omega_q$ one can often make an additional \emph{secular approximation} which we discuss in detail later on. Breaking the assumption that the qubit's level splitting is the largest energy leads to the ultrastrong coupling regime, which is studied elsewhere\cite{Ashhab2010}.

We model the time evolution of the system's reduced density matrix $\rho$ using the Bloch-Redfield equations\cite{Bloch1957}$^,$ \cite{Redfield}. Using the eigenvectors $\ket{1}$ to $\ket{8}$ (given in appendix \ref{app eigenstates}) of $H_{sys}$ as basis states, the Bloch-Redfield equations read element-wise:
\begin{align}
	\dot \rho_{nm} = -i \omega_{nm} \rho_{nm} + \sum_{n' m'} \mathcal{R}_{nmn'm'} \rho_{n'm'} \, \label{bloch-redfield}
\end{align}
with the Redfield tensor:
\begin{align}%*}
	\begin{aligned}
	\mathcal{R}_{nmn'm'} &:= \Lambda_{m'mnn'}+ \tilde \Lambda_{nn'm'm} - \sum_k (\Lambda_{nkkn'} \delta_{mm'} + \tilde \Lambda_{kmm'k}\delta_{nn'}) \,\\
	\Lambda_{nmn'm'} &:= \hat s_{nm} \hat s_{n'm'} \frac{1}{2} C(\omega= \omega_{m'n'}) \,\\
	\tilde \Lambda_{nmn'm'} &:= \hat s_{nm} \hat s_{n'm'} \frac{1}{2} C(\omega= \omega_{n'm'}) \,. 
	\end{aligned}
\end{align}%*}
Here $\rho_{nm} = \bra{n} \rho \ket{m}$ denotes the density matrix element at position $n,m$ and $\omega_{nm}:= \omega_n - \omega_m$ is the energy difference of the Hamiltonian eigenstates $\ket{n}$ and $\ket{m}$ of the system. The system operator which couples to the environment $\hat s$, is defined in $H_{int}$. In this approach the environment is solely characterized through its spectral function
\begin{align}
	&C(\omega):= \int_{-\infty}^\infty d \tau \; e^{i \omega \tau} \, \langle  \tilde B(\tau) \tilde B(0) \rangle \label{spectral function}
\end{align}
where the bath operator $\tilde B$ is taken in the interaction picture defined with respect to $H_{int}$. This spectral function, eq. \ref{spectral function}, is assumed to change slowly such that it does not change on the small scale of $\delta$ and $g$ (but only on the much larger scale of $2\omega_q$). 

The transversal coupling to the low temperature bath leads to unidirectional population transfers to states with lower excitation numbers, i.e.~relaxation. These transfers appear in the Bloch-Redfield equations as linear dependencies of the time-derivatives of certain diagonal elements of the system's density matrix on other diagonal elements, each with a coefficient. These coefficients (i.e.~relaxation transition rates) are all proportional to $v_\perp^2 C(2 \omega_q)$. For later use we define a general relaxation rate due to coupling of the TLS to the environment:
\begin{align}
\Gamma_1 &:= v_\perp^2 C(2 \omega_q) \, \label{relaxation rate definition}
\end{align}

The longitudinal bath coupling leads to two processes: First a loss of phase coherence between the states of the system, i.e.~dephasing, which is mathematically represented by the decay of off-diagonal elements in the density matrix. Second a mutual population transfer between certain eigenstates\cite{Boissoneault2009} with the same excitation number: $|2\rangle, |4\rangle$ and $|5\rangle, |7\rangle$ (figure \ref{fig system}\textbf{C}). The corresponding decay rates and transition rates are all similarly proportional to:
\begin{align}
\Gamma_\varphi &:= v_\parallel^2 C(0) \, \label{dephasing rate definition}
\end{align}
An energy diagram of the eigenstates is depicted in figure \ref{fig system}\textbf{C} and all transitions are shown as arrows.

When the TLS is decoupled from the qubits ($g_1=g_2=0$) then $\Gamma_1$ is the decay rate of the population of its excited state and $2 \Gamma_\varphi$ is the additional decay rate of its two off-diagonal elements, i.e.~its relaxation and pure dephasing rate respectively. %Bloch-Redfield-Einstieg.nb

Comparing the resulting Bloch-Redfield equations with an approach assuming Lindblad equations\cite{Breuerbook, Lindblad1976} with a phenomenological relaxation rate and dephasing rate on the TLS we find that the two sets of differential equations are equivalent when the following two conditions are met. First the spectral function should not change on the scale of $\delta$ and $g$. As the second condition one of the following three requirements has to be fulfilled: Either (i) the full secular approximation (explained in the next paragraph) is applied to both the Lindblad and Bloch-Redfield equations or (ii) we take only longitudinal TLS-bath coupling, i.e., $v_{\perp} = 0$ or (iii) we assume only transversal TLS-bath coupling, $v_\parallel = 0$, and choose an initial state which is confined to the single excitation subspace. In case of equivalence the two phenomenological rates in the Lindblad equations can be identified as our definitions $\Gamma_1$ and $\Gamma_\varphi$.

The full secular approximation neglects all dependencies between different elements of the system's density matrix if at least one of them is an off-diagonal element. The necessary and sufficient condition for this approximation is that the system's level splittings and their differences are large compared to the decoherence rates. Physically this means assuming $\omega_q \gg  g \gg \Gamma_1, \Gamma_\varphi$ and $\omega_q \gg |\delta|$ i.e.~the TLS is somewhat coherent. %"B-R 2q1t detuned include c31 also q2 and f.nb"

In the following sections, analytical solutions to the Bloch-Redfield equations in different regimes are discussed. For  $\delta=0$ these solutions are given in appendix \ref{app Fourier transform}. The first solution we show (appendix \ref{app weak decoherence solutions}) is obtained using the full secular approximation and is valid for what we call the weak decoherence regime, when the resulting decoherence rates are smaller than the coupling strength $g$ between the qubits and the TLS. We obtain two further analytical solutions for purely transversal (appendix \ref{app purely transversal coupling solutions}) i.e.~$v_\parallel=0$, and purely longitudinal (appendix \ref{app purely longitudinal coupling solutions}) i.e.~$v_\perp = 0$, TLS-environment coupling. The combination of our particular initial state, the large $\omega_q$ limit and $v_\perp v_\parallel=0$ allows us to solve the master equation without the secular approximation. That means no assumption about the relative sizes of $g$ and $\Gamma_1, \Gamma_2$ has to be made. These last two solutions are therefore also valid for strong decoherence, i.e., when the decoherence rates are bigger than the coupling strength $g$. 

\section{Dynamics \label{sec general physics}}
In this section we present the analytical solutions to the Bloch-Redfield equations for our system. We start the section by summarizing the results for a single qubit coupled to a TLS for later comparison.  Following this, in section \ref{two qubit} we study the behaviour of two qubits coupled to a TLS in detail.

\subsection{A single qubit coupled to a TLS \label{one qubit}}
\begin{figure*}
\includegraphics[scale=1]{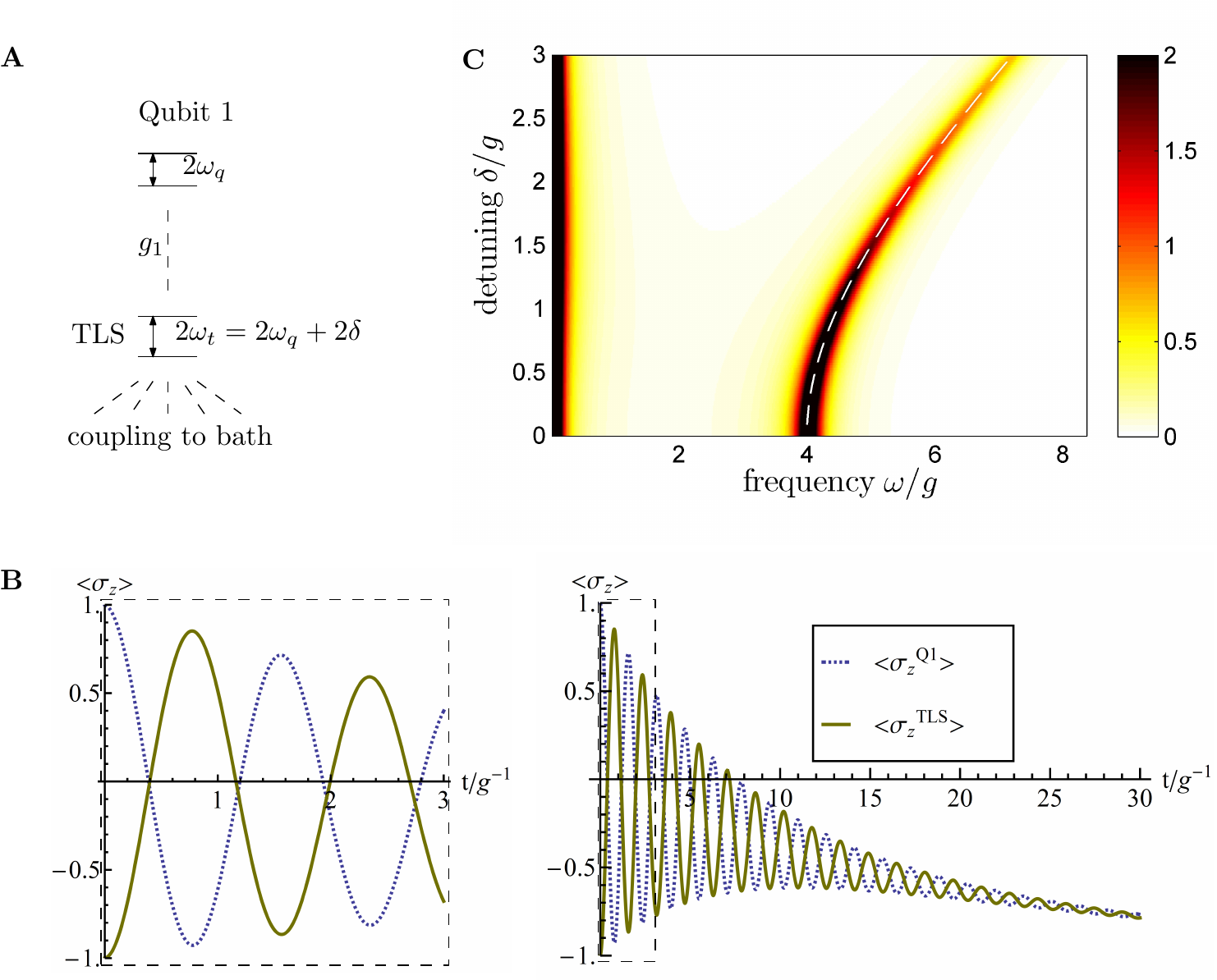}
	\caption{\textbf{A} Illustration of the simplified model system of one qubit with level splitting $\omega_q$ coupled with coupling strength $g_1$ to an environmental TLS detuned by $\delta$. The TLS in turn is coupled to a bath. This system follows from figure \ref{fig system}\textbf{A} by setting $g_2=0$. \;\textbf{B} Expectation values for the case of a single qubit coupled to a TLS as a function of time for $\Gamma_1=\Gamma_\varphi=0.1 g_1; \; g_2=0; \; \delta=0$ \;\textbf{C} Real part of the one-sided Fourier transform of $\langle \sigma_z^{Q1} \rangle(t)$ as a function of detuning $\delta$ with $\Gamma_1=\Gamma_\varphi = 0.1 g_1; \; g_2=0$. This is a numerical solution of either the Bloch-Redfield equations, where $\Gamma_1$ and $\Gamma_\varphi$ are the definitions given by eq.~\ref{relaxation rate definition} and \ref{dephasing rate definition} or a numerical solution of the Lindblad equations with phenomenological rates.
	Analytically we find the angular frequency of the oscillation: $\omega_{osc}=2\sqrt{\delta^2+4g_1^2}$ which is drawn as a dashed line. For details of the calculation, see section \ref{one qubit}. 
	\label{fig 1qubit_figure}
	} 
\end{figure*}

Before we study the more complicated case of two qubits, the dynamics of a single qubit coupled to an environmental TLS (see figure \ref{fig 1qubit_figure}\textbf{A}) provides a clear overview of the relevant physics. This special case can be obtained from all solutions by setting $g_2=0$ and tracing out the second qubit. Performing this on the Hamiltonian, eq.~\eqref{system Hamiltonian}, yields (for corresponding eigenstates see appendix \ref{app eigenstates}):
\begin{align}
H_{1q} &= \omega_q \sigma_z^{Q1} +  (\omega_q + \delta )\sigma_z^{TLS} + g_1 (\sigma_x^{Q1} \sigma_x^{TLS} + \sigma_y^{Q1} \sigma_y^{TLS}) 
\end{align}

A comparison with the results in this section will later allow us to distinguish phenomena which depend purely on the existence of two qubits and those which are due to qubit-TLS coupling in general. This section strongly depends on previous work \cite{Mueller2009}, which is reproduced in our notation. As observable we consider the expectation value $\langle \sigma_z^{Q1} \rangle $ (which is proportional to the qubit's energy). Equivalently, one could use the probability to find the qubit in the excited state, $P_{\text{exc}}=\frac{1}{2} \left( \langle \sigma_z^{Q1} \rangle + 1 \right)$.

Assuming weak decoherence (i.e.~we take the full secular approximation in the Bloch-Redfield equations) and no detuning $\delta=0$ one finds the expectation values as a function of time: 
\begin{align}
\langle \sigma_z^{Q1} \rangle(t) &= 
-1 + e^{- \frac{\Gamma_1}{2} t} + e^{(-\frac{\Gamma_1}{2}-\Gamma_\varphi) t}  \cos \left[4 g_1 \,t\right] \\
\langle \sigma_z^{TLS} \rangle(t) &= 
-1 + e^{- \frac{\Gamma_1}{2} t} -  e^{(-\frac{\Gamma_1}{2}-\Gamma_\varphi) t} \cos\left[4 g_1 \;t\right]
\end{align}
The population oscillates between the qubit and the TLS  (cf. Fig \ref{fig 1qubit_figure}\textbf{B}) with the oscillation frequency proportional to their transversal coupling strength. The oscillations decay on the time scale corresponding to the decoherence rates of the TLS. 

Equivalently we can consider the evolution in Fourier space, where the frequency and the decay rate are equal to  the position and width respectively of the corresponding frequency peak (for details see appendix \ref{app Fourier transform 4}). 
In figure \ref{fig 1qubit_figure}\textbf{C} the real part of the one-sided Fourier transform of $\langle \sigma_z^{Q1} \rangle(t)$ is plotted as a function of detuning between the qubit and the TLS. The peak which starts at frequency $4 g_1$ diminishes with increasing detuning, indicating a shift from oscillatory behaviour to pure exponential decay. The analytical solution for $\langle \sigma_z^{Q1} \rangle (t)$ with $\delta \neq 0$ contains complicated coefficients\cite{Mueller2009}, but the frequency of the oscillation is simply $\omega_{osc}=2 \sqrt{\delta^2+4g_1^2}$. This corresponds to the level splitting between the hybridized states $\ket{2}_{1q}$ and $\ket{4}_{1q}$ (given in appendix \ref{app eigenstates}) and is plotted as a dashed line in figure \ref{fig 1qubit_figure}\textbf{C}. The oscillatory behaviour is described by this one frequency, which corresponds to the standard ``generalized Rabi frequency''\cite{Scully1999} from quantum optics. For large detuning $\delta$ the expectation value $\langle \sigma_z^{Q1} \rangle(t)$ is dominated by one purely decaying term. A Taylor expansion for small $g/\delta$ on the corresponding decay rate shows that the rate vanishes with increasing $\delta/g$ as $ (\Gamma_1+4\Gamma_\varphi)g^2/\delta^2 \rightarrow 0$.%Taylor expansion in B-R 1q1t detuned include c31.nb

So far only weak decoherence on the TLS was considered. We also want to consider the limit where the decoherence is stronger than the qubit-TLS coupling. Simplifying the equations to purely transversal bath coupling ($v_\parallel = 0 \Rightarrow \Gamma_\varphi =0$) one finds analytical solutions for the system dynamics without the use of the secular approximation and therefore valid for stronger decoherence:
\begin{multline}
\langle \sigma_z^{Q1} \rangle(t) = -1 + \frac{-64 g_1^2}{\mu ^2} e^{-\frac{t \Gamma_1}{2}} + 2 e^{-\frac{t \Gamma_1}{2}} 
\left(\frac{-32 g_1^2+\Gamma_1^2}{\mu ^2} \cosh \left[\frac{t \mu }{2}\right]+ 
\frac{\Gamma_1}{\mu } \sinh \left[\frac{t \mu }{2}\right] \right) \label{energy for single qubit} 
\end{multline}
where $\mu := \sqrt{\Gamma_1^2 - 64 g_1^2}$.
From this expression we see that as the decoherence rate $\Gamma_1$ increases relative to the qubit-TLS coupling strength $g_1$, the dynamics changes from oscillations to pure exponential decay. This becomes obvious by rewriting the hyperbolic cosine as:
\begin{align}
\cosh \left(\frac{1}{2} \sqrt{\Gamma_1^2-64 g_1^2} \;t \right) = 
 \left\{ \begin{aligned} \cos\left(\frac{1}{2} \sqrt{64 g_1^2-\Gamma_1^2} \;t\right) \quad \text{ for }8g_1 > \Gamma_1 \\ 
\frac{1}{2}  \left(e^{+ ...}+e^{- ...}\right) \quad \text{ for } 8g_1 < \Gamma_1\end{aligned} \right. \label{rewrite cosh}
\end{align}
and similarly for the hyperbolic sine functions. Therefore, we can identify the threshold between oscillations and decay in our approximations as precisely $\Gamma_1=8g_1$.

\begin{figure*}
\includegraphics{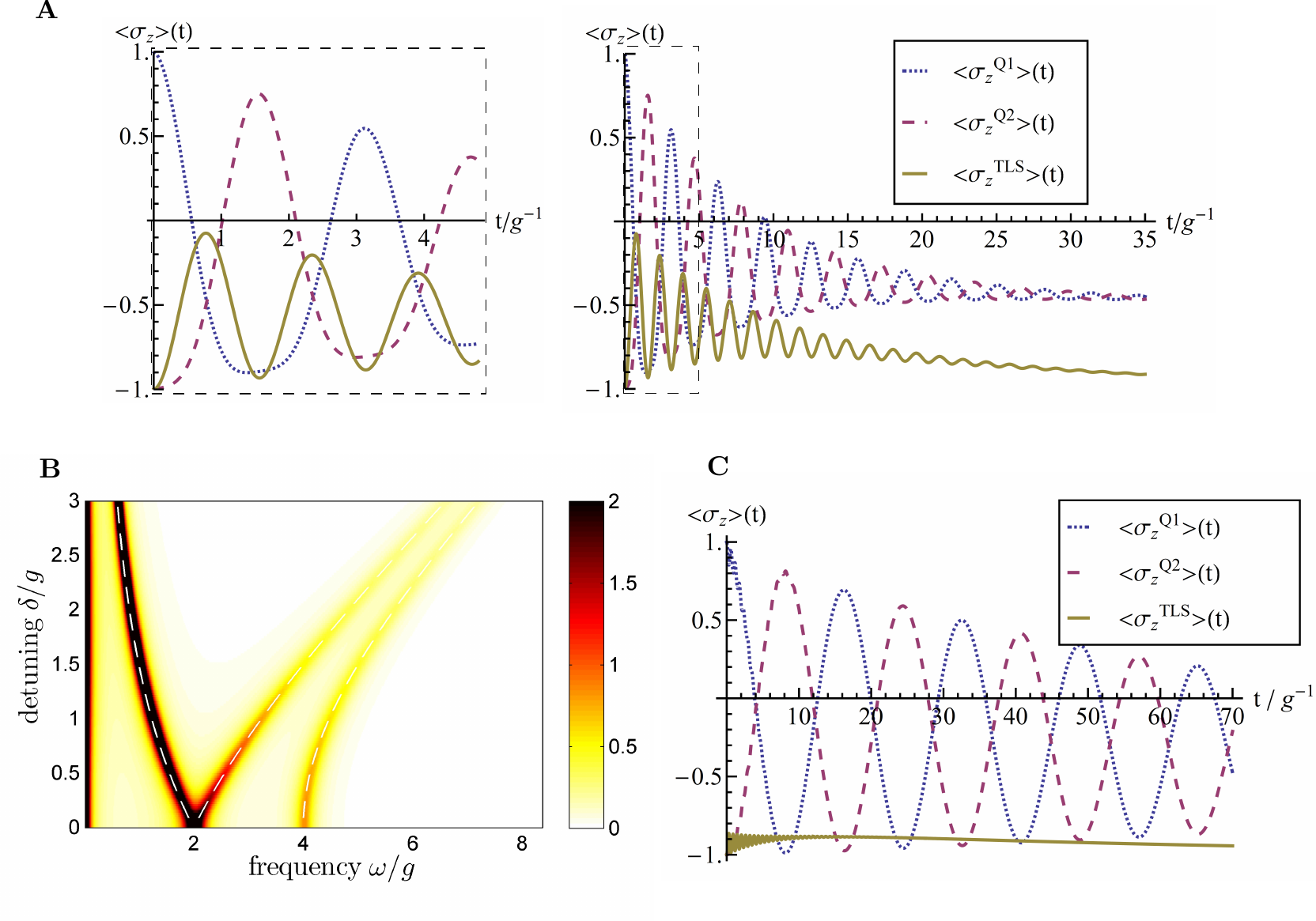}
\caption{\textbf{A} Expectation values of both qubits and the TLS. The excitation is shifted from one qubit via the TLS to the other qubit and back. The parameters are chosen as: $\Gamma_1=\Gamma_\varphi=0.1 g;\; g_1=g_2;\; \delta=0$ \; \textbf{B} Real part of the one-sided Fourier transform of $\langle \sigma_z^{Q1} \rangle(t)$ as a function of detuning $\delta$. Analytically we find the three oscillation frequencies $- \delta + \sqrt{\delta^2 + 4 g^2} \,,\quad \delta + \sqrt{\delta^2 + 4 g^2} \,,\quad 2 \sqrt{\delta^2 + 4 g^2}$ which are given by the dashed lines in the plot. The parameters are: $\Gamma_1=\Gamma_\varphi=0.1 g;\; g_1=g_2$ \; \textbf{C} Time evolution of the expectation values for strong detuning $\delta = 5g;\; \Gamma_1=\Gamma_\varphi=0.1g;\; g_1=g_2$. Even though the TLS is only minimally excited, the effective coupling mediated by it still leads to coherent exchange of energy between the two qubits.
\label{fig 2qubit_energies}}
\end{figure*}

\subsection{Two qubits coupled to a TLS \label{two qubit}}
Having reviewed the behaviour of a single qubit coupling to an environmental TLS, we now consider the behaviour of a dual-probe configuration. Such a system is of particular interest when there is no direct coupling between the qubits. This situation allows us to probe what we call \emph{coherent pockets} of the environment. When such a pocket is present in the environment, probing simultaneously with two qubits shows qualitatively different behaviour to the standard weakly coupled, Markovian environment, which would affect each qubit independently.

\subsubsection{Mediated coupling between the qubits in the weak decoherence regime}
We will first show how the coupling between qubits and TLS will mediate an effective interaction between the two qubits themselves. For simplicity we initially consider $\delta=0$, i.e.~both qubits are resonant with the TLS. In this case, and for the initial state chosen in section \ref{model} the energy from qubit 1 coherently oscillates between the two qubits via excitation of the TLS (figure \ref{fig 2qubit_energies}\textbf{A}). In contrast to the simpler case presented in section \ref{one qubit}, the oscillations now show two distinct frequencies, namely $4g$ and $2g$. The smaller of the two frequencies corresponds to the oscillation of energy between the two qubits. The full analytical expression for $\mean{\sigma_{z}^{Q1}}$ can be found in appendix \ref{app Fourier transform}. 

Again the change in the system dynamics due to detuning $\delta \neq 0$ can be understood best by regarding the Fourier transform of the expectation value $\langle \sigma_z^{Q1} \rangle (t)$. This yields a peak for each term located at the corresponding frequency whose half-width-at-half-maximum (HWHM) equals the decay rate of the corresponding oscillatory component. Figure \ref{fig 2qubit_energies}\textbf{B} shows the result of a numerical solution of the Bloch-Redfield equations with dashed lines indicating the analytical expressions for the frequency shifts due to the detuning. With increasing detuning $\delta$, the peak at frequency $2g$ splits into two different frequency peaks. The amplitude of the two high frequency contributions diminishes with stronger detuning $\delta$, while the amplitude of the lower frequency peak increases. This means for stronger detuning $\delta > g$, the energy oscillates between the qubits mainly at the lower frequency $\omega_{low}=|\delta| - \sqrt{\delta^2 - 4 g^2}$. For sufficiently strong detuning, the TLS is largely unpopulated during this process (figure \ref{fig 2qubit_energies}\textbf{C}). 

This kind of off-resonant interaction with the TLS leads to an effective transversal coupling between the qubits.  This is the usual dispersive coupling term~\cite{gerrybook, Raimond2001, Sorensen1999, Zheng2000, Majer2007}, in this case due to virtual excitation of the TLS.  Performing a Taylor expansion for $g/\delta \ll 1$ on both the lower oscillation frequency $\omega_{low}$ and on the decay rate of this oscillating term in the weak decoherence solution with detuning yields $\omega_{low} \approx 2 g^2/|\delta|$ and $\gamma_{low} \approx g^2 (\Gamma_1 + 12 \Gamma_\varphi) / 6 \delta^2$. The lower frequency can here be interpreted as the effective coupling strength between the qubits $\omega_{low}=g_{eff}$. This effective coupling approaches zero slower than the decay rate $\gamma_{low}$ as the magnitude of the detuning increases. The strongly detuned TLS therefore mediates an effective transversal coupling between the qubits with a weakened influence of the TLS' decoherence rates. However, ultimately the effective coupling strength (i.e.~the frequency of the oscillation) approaches zero for $\delta / g \rightarrow \infty$. % Taylor expansion in Diplomarbeit, \gamma_{low} in B-R 2q1t detuned include c31.nb

Here we see a fundamentally different behaviour as compared to a single qubit coupled to a TLS. The oscillations do not change to a pure decay for strong (but not yet infinite) detuning $\delta$. Additionally the frequency of these oscillations approaches zero much slower ($\propto g/\delta$) than the decay rate of the single qubit ($\propto g^2/\delta^2$). 
% find proportionality for taylor series of \Gamma_{34} in B-R 1q1t detuned include c31.nb (other terms are zero for large detuning).

This result has important implications for future experimental designs involving several qubits in a closely confined space. There the occurrence of an environmental TLS, which couples to two qubits at once might have a realistic probability, especially in solid-state qubits. In that case the qubits are affected by the TLS over a wide range of detuning, causing effective coupling between the qubits. 

\subsubsection{Formation of stray entanglement}
As we can see in Fig 3A, the steady state of both qubits is not their respective ground state. Rather, they decay into a state with a finite probability of finding them excited. This behaviour can be attributed to the existence of a so called dark state in our system.
The state $\ket{3} = \frac{g_2}{g} \ket{ \uparrow \downarrow \downarrow } - \frac{g_1}{g} \ket{ \downarrow \uparrow \downarrow }$ (appendix \ref{app eigenstates}) is an entangled state of both qubits with the TLS in its ground state. 
The amplitudes of the two states (with the respective qubit excited) have a relative complex phase of $\pi$ in the time evolution which leads to a cancellation of the qubits' influence on the TLS. In our system, this state is thus not influenced by decoherence and the system will remain in it for a long time (i.e. for the intrinsic decoherence time of the qubits). This is simply a manifestation of the physics of super- and sub-radiance\cite{Crubellier1985} due to the interfering pathways from the qubits to the TLS. Since our chosen initial state is a statistical mixture including the eigenstate $\ket{3}$, the steady state of the system will still include this fraction of the dark state. 
The entanglement of the two qubits in the steady state depends on the interplay of two things: the ``concurrence'' \cite{Wootters98} of the dark state by itself which is given by $C=2g_1 g_2 / g^2$ (i.e.~a Bell state for $g_1=g_2$) and the fraction of the dark state in the mixed steady state. Taking both into consideration we find the maximal ``entanglement of formation''\cite{Wootters98} of the final state as $E = 53\%$ which is reached for $g_1=g_2/\sqrt{3}$.

\subsubsection{Threshold between weak and strong decoherence}
\begin{figure}
\centering
\includegraphics[width=0.5\columnwidth]{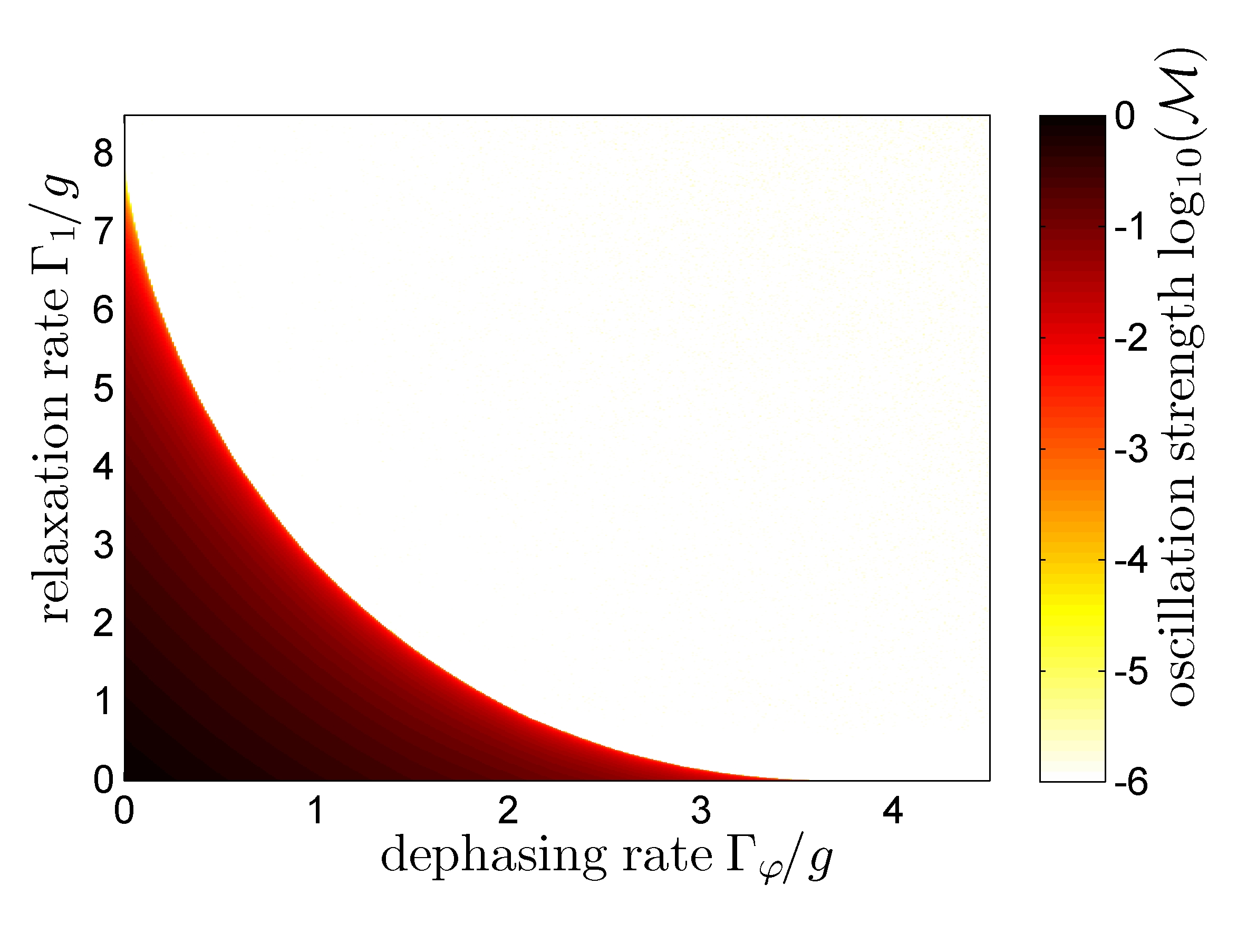} 
\caption{$\text{Log}_{10}$ of the oscillation strength $\mathcal{M}$ (eq. \ref{oscillation strength}) as a function of the decoherence rates $\Gamma_1$ and $\Gamma_\varphi$. The threshold between oscillations and decay can be seen as a drop in the oscillation strength by four orders of magnitude. We solved the full Bloch-Redfield equations numerically (without the secular approximation) and for simplicity we set $g_1=g_2$ and $ \delta=0$. The level splitting was $\omega_q=1000 g$.
\label{fig decayosc_threshold} }
\end{figure}

When the decoherence rates of the TLS are stronger than the qubits-TLS coupling $\Gamma_1, \Gamma_\varphi > g$ the secular approximation (section \ref{model}) can no longer be fully applied. With increasing decoherence rates the dynamics of the three subsystems changes from an oscillating behaviour to a pure decay. 
This behaviour is analogous to a single qubit coupled to a strongly decoherent TLS. In section \ref{one qubit} we saw that in this case the crossover was defined by the point $\Gamma_1 = 8 g_1$. For two qubits, the crossover between the weak and strong decoherence regimes is investigated numerically. For oscillations to occur between the qubits and the environmental TLS there needs to be an instant in time in which the population of the TLS is larger than both qubits combined. We therefore define the maximum value in the evolution: 
\begin{align}
\mathcal{M}= \max_t \quad \left\{ \langle \sigma_z^{TLS} \rangle (t) - \left[\langle \sigma_z^{Q1} \rangle (t) + \langle \sigma_z^{Q2} \rangle (t) \right] \right\} \label{oscillation strength}
\end{align}
as a measure of the strength of the oscillation. In the regime of strong decoherence the energy of the qubits decays via the TLS to the environment and our defined measure is always zero. In the weak decoherence (i.e., oscillating) regime the energy leaves the qubits and then partially returns via coherent oscillations from the TLS. This gives a positive value for the defined measure. Figure \ref{fig decayosc_threshold} is a logarithmic plot of this measure as a function of the two decoherence rates $\Gamma_1 $ and $\Gamma_\varphi$. There is a sudden drop in the oscillation strength to negligible values marking a clear threshold between the oscillating (weak decoherence) and the decaying (strong decoherence) regime. The oscillating regime (dark area) also marks precisely the parameter regime in which the full secular approximation is valid.

For purely transversal (respectively purely longitudinal) TLS-bath coupling $v_\parallel=0 \Rightarrow \Gamma_\varphi =0$ (respectively $v_\perp = 0 \Rightarrow \Gamma_1=0$) an analytical solution can be found. Analogous to eq. \eqref{energy for single qubit} and \eqref{rewrite cosh} we find the analytical threshold between strong and weak decoherence regime at the two points:
\begin{align}
\begin{aligned}
\Gamma_1&=8g, \Gamma_\varphi=0 \\
&\text{and} \\
\Gamma_\varphi&= 4g, \Gamma_1=0 
\end{aligned}\label{exact threshold}
\end{align}
This corresponds to the point where the threshold in figure \ref{fig decayosc_threshold} crosses the two axes. The factor of two between $\Gamma_1$ and $\Gamma_\varphi$ stems directly from the definition of the rates (eq.  \ref{relaxation rate definition} and \ref{dephasing rate definition}) as they appear in the master equations i.e. the off-diagonal elements of the uncoupled TLS-density matrix decay with a rate $2\Gamma_\varphi$). 

\subsubsection{Markovianity}
\begin{figure}
\centering
\includegraphics[width=0.25\columnwidth]{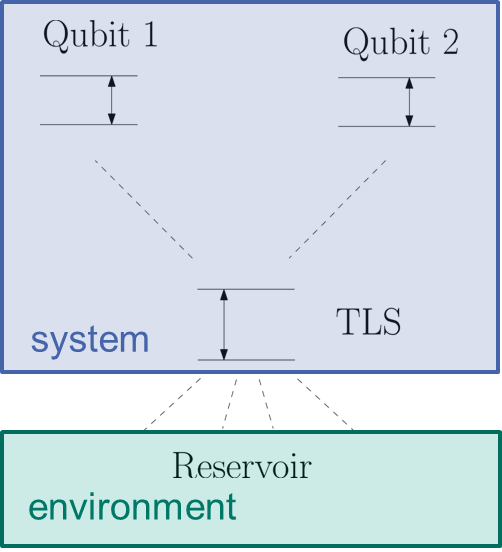} \qquad 
\includegraphics[width=0.25\columnwidth]{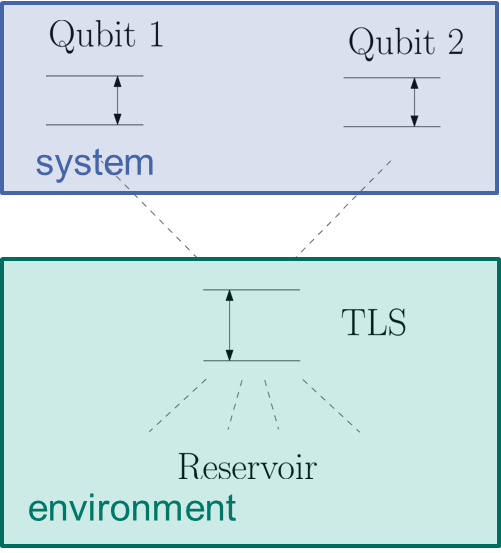}
\caption{Two possibilities of placing the system/environment boundary: The left case was used to set up and solve the Bloch-Redfield equations for which the system dynamics is by default assumed to be Markovian. For the configuration on the right side, where the TLS is seen as a part of the environment, we investigate Markovianity of the system of two qubits, see text.
\label{fig systemenvironmentdivision}}
\end{figure}

The coherent coupling to environmental states usually leads to non-Markovian\cite{Breuerbook, Cui2008} dynamics in the system (excluding the environmental states). Using our model, we can choose where to draw the system/environment boundary (see figure \ref{fig systemenvironmentdivision}) and therefore explore this behaviour in a systematic fashion. Regarding the TLS as part of the system, Markovian dynamics is assumed by default as this is a necessary condition to apply the Bloch-Redfield equations. However, tracing out the TLS we investigate Markovianity of the two qubit system.

Several measures exist to theoretically quantify non-Markovianity\cite{Breuer2009, Rivas2010, Lu2010} in qubit systems. For our purposes we use the connection between the degree of non-Markovianity and an increase in the time evolution of the distinguishability, calculated with the trace-distance\cite{Breuerbook} $D(t)=Tr\left\{\sqrt{[\rho_1(t)-\rho_2(t)]^2}\right\}$ of two different initial states. By tracing out the environmental TLS in the analytical solution for purely transversal TLS-bath coupling (thus reducing the density matrix to the system of the two qubits) one can find the trace-distance between the two time evolutions of the initial states $|Q1, Q2\rangle = \ket{ \uparrow \downarrow }$ and $|Q1, Q2\rangle =\ket{ \downarrow \uparrow \rangle }$. We find that the trace-distance only increases in the weak decoherence (oscillating) regime. This means that it is the same analytically exact threshold between oscillating behaviour and decaying behaviour (compare eq. \eqref{exact threshold}) that divides non-Markovian from Markovian behaviour. The chronologically first and strongest increase $\Delta D_{\uparrow}$ in the trace-distance is
\begin{align}%*}
\Delta D_{\uparrow} = \left\{ \begin{aligned} & \exp\left( -\frac{\pi  \Gamma _1}{\sqrt{64 g^2-\Gamma _1^2}}\right) \quad &\text{ for }8g \geq \Gamma_1 \\
&0 \quad &\text{ for } 8g < \Gamma_1\end{aligned} \right.
\end{align}%*} %file: markovianity trace-distance Lindblad.nb
for the particular case of pure relaxation and equal qubit-TLS couplings ($\Gamma_\varphi = 0,\; g_1=g_2=g/\sqrt{2} $), where $\Delta D_{\uparrow} =0$ corresponds to a monotonically decreasing trace-distance, i.e.~no increase in $D(t)$.
Here we see that our experimentally measurable definition of the oscillation strength $\mathcal{M}$ is in this case equivalent to a measure of non-Markovianity. We can therefore see the direct link between probing coherence in the environment and probing non-Markovianity.

\subsubsection{Effective decay rate}
In order to characterize the decohering influence of the TLS on our probe qubits, we want to introduce a single, \emph{effective} decay rate, serving as a figure of merit towards determining the influence of the TLS' decoherence rates $\Gamma_1, \Gamma_\varphi$. In general, the time evolution of the qubits' expectation values are given by sums of exponential functions, where, for the weak decoherence regime, some of the terms will be decaying oscillations. In the strong decoherence regime, the effective decay rate is introduced by replacing the sum of exponentials $f(t)=\sum_j c_j \exp (-\gamma_j t)$ by a single exponential $c \exp(-\gamma_{eff} t)$ with the two conditions:
\begin{align}%*}
1) &\int_0^\infty dt f(t) = \int_0^\infty dt \,c \exp(-\gamma_{eff} t)\\
2) &\,f(0) = c
\end{align}%*}
which leads to the simple formula:
\begin{align}
\gamma_{eff}=\frac{\sum_j c_j}{\sum_j c_j/ \gamma_j} \label{effective decay rate calculation}
\end{align}
This defines a single decay rate in the strong decoherence regime. From the two analytical solutions which are valid for strong decoherence given in appendix \ref{app Fourier transform} (zero detuning and either purely longitudinal or purely transversal TLS-bath coupling) we see that
\begin{align}
\gamma_{eff, \perp} &= \frac{16 g^2 \left(g_1^2+2 g_2^2\right) \Gamma _1}{16 g^2 g_1^2+\left(g_1^2+4 g_2^2\right) \Gamma _1^2}  & \text{ for } \Gamma_1>8g, \Gamma_\varphi=0 \label{effective decay rate transv}\\
\gamma_{eff, \parallel} &= \frac{8 g^2 \left(g_1^2+4 g_2^2\right)}{\left(g_1^2+16 g_2^2\right) \Gamma _2} & \text{ for } \Gamma_\varphi > 4g, \Gamma_1=0 \label{effective decay rate longi}
\end{align}

\begin{figure}
\centering
\includegraphics[width=0.5\columnwidth]{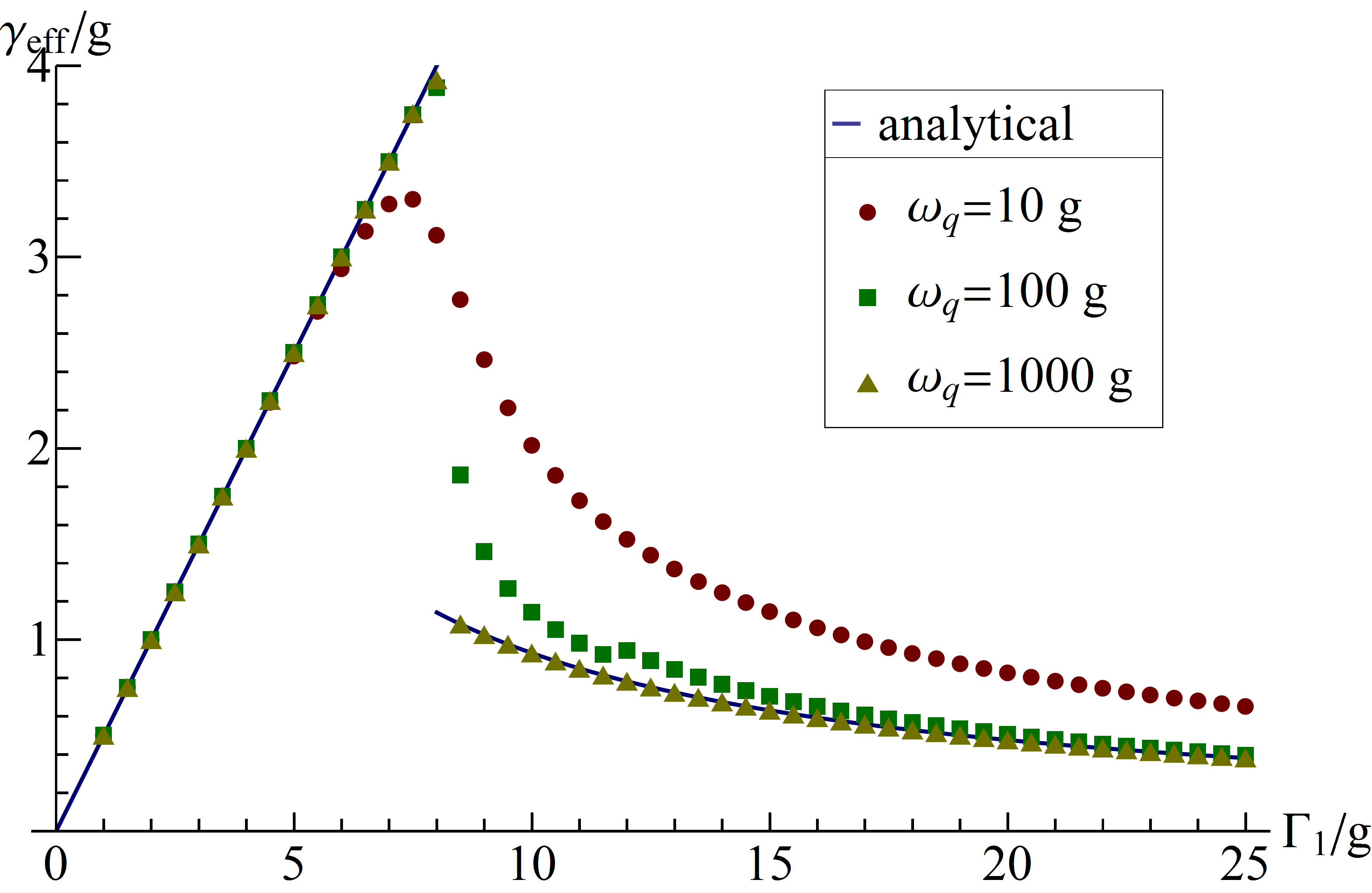}
\caption{Effective decay rate $\gamma_{eff}$ of the energy of qubit 1 as a function of the TLS relaxation rate $\Gamma_1$ (both rates in units of $g$). In the weak decoherence regime they are linearly dependent (eq. \ref{effective decay for weak dec}), for strong decoherence the effective decay rate decreases with increasing relaxation rate $\Gamma_1$ (eq. \ref{effective decay rate transv}). The other parameters of the plot are: $g_1=g_2, \Gamma_\varphi=0$ \label{fig effectivedecayrate} }
\end{figure}

For purely transversal bath coupling the effective decay rate $\gamma_{eff, \perp}$ is plotted in figure \ref{fig effectivedecayrate}. Interestingly the effective decay rate eq. \eqref{effective decay rate transv} and \eqref{effective decay rate longi} monotonically decreases with decreasing ratio $g /\Gamma_1$ ($g/\Gamma_\varphi$ respectively).  The observation of \emph{weaker} decoherence on the qubit with \emph{stronger} decoherence rate of the TLS is due to ``blocking'' of dynamics of the TLS due to strong decoherence: The exchange of energy between the qubits and the TLS is slowed down and thereby also the loss of energy to the environment. Regarding decoherence as a measurement process this is analogous to the Zeno-effect\cite{Breuerbook}. 

This behaviour is in contrast to the weak decoherence regime, where both purely decaying exponents and decaying oscillations occur. We find two ways of defining an effective decay rate: either describe the decay of the envelope of the oscillations or the effective decay of their average. For details, we refer the reader to Ref.\cite{Mueller2009} and appendix \ref{app average and envelope}. Here we use the decay of the average and find the effective decay rate $\gamma_{eff}$ to be linearly dependent on the relaxation rate $\Gamma_{1}$:
\begin{align}
\gamma_{eff}&=\frac{1}{2} \Gamma_1 & \text{ for } \Gamma_1<8g, \Gamma_\varphi<4g \label{effective decay for weak dec}
\end{align} 
In the intermediate regime $2 \lesssim \Gamma_1/g < 8$ and $1 \lesssim \Gamma_\varphi /g < 4$, where the lower bound is found empirically, the oscillations are slow on the timescale of the decay. In this case the first half oscillation, transmission of energy from the qubit into the TLS, dominates the behaviour, and the average decay rate does not reproduce the behaviour well. An effective decay rate is not a good description of the dynamics in this regime and the apparent discontinuity in figure \ref{fig effectivedecayrate} actually appears as a smooth transition in the time evolution of the qubits' expectation value. We have also plotted three numerical calculations in figure \ref{fig effectivedecayrate} for different level splittings $\omega_q$ (while $\delta$ is always zero). As stated in section \ref{model} the analytical result is obtained in the single excitation subspace which requires $\omega_q \gg g_1, g_2,\delta$. We see excellent agreement between the analytical solution and the numerical solution of the full Bloch-Redfield equations for $\omega_q \gtrsim 1000 g$.

\section{Probing a single TLS with two qubits: Parameter extraction \label{sec probing}}
After studying the system and its dynamics in the previous sections, we now interpret the results in the context of decoherence microscopy. In particular we focus on the ability to obtain the TLS parameters with a dual probe and compare it with a single qubit-probe.  For this purpose we only consider the Fourier transform of the evolution of the qubits' excited-state-population, as this conveniently represents the parameters of interest.

\begin{figure}
\centering
\includegraphics[width=0.5 \columnwidth]{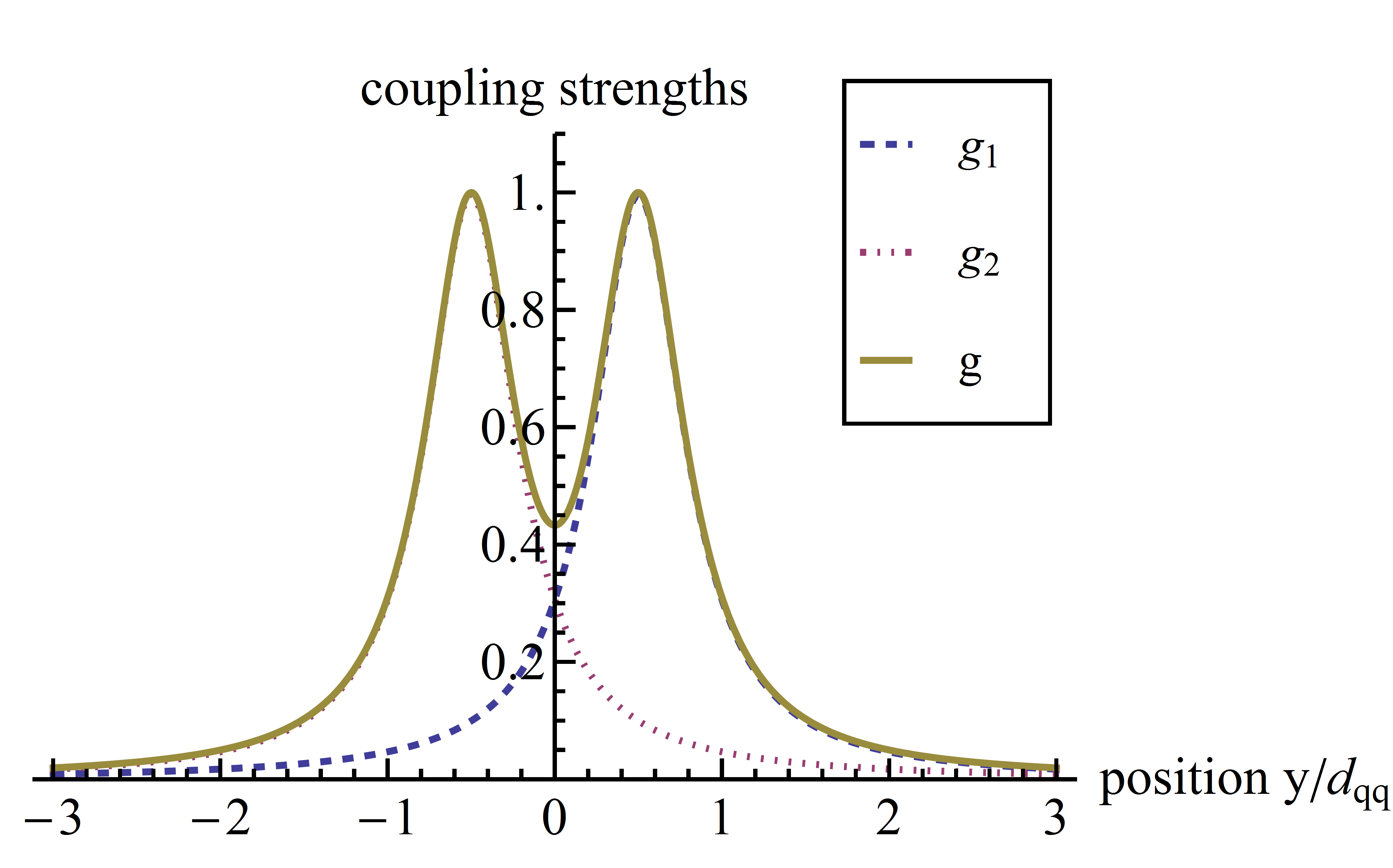}
\caption{Characteristic behaviour of the coupling strengths to the TLS in the substrate as a function of the position of the two qubits $y$ (compare figure \ref{fig system}\textbf{A}). The distance between the peaks is controlled by the distance between the qubits $d_{qq}$. The width of the peaks is controlled by the height $h$ above the sample. Here we chose $d_{qq}=3h$. The coupling strengths are normalized such that the maximum value is 1. \label{fig couplingplot}}
\end{figure}

Now we will give a more concrete form to the theoretical coupling parameters $g_1$ and $g_2$ from eq.\ref{system Hamiltonian}.  If we assume the coupling strengths depend on the distances $d_1$ and $d_2$ between the qubits and the TLS, as $g_j \propto  1/ d_j^2$ and the qubits are moved along their connecting line above the TLS in the substrate (see figure \ref{fig system}) then the coupling strengths behave characteristically as a function of the position $y$ (figure \ref{fig couplingplot}). 

\subsection{Weak decoherence regime - oscillating behaviour}
From an experimental point of view the parameters $\delta, \Gamma_1, \Gamma_\varphi$ of the environmental TLS and even the coupling strengths $g_1, g_2$ are in general unknown. We first consider the weak decoherence regime when the qubits are close enough to the TLS (so that $g \gg \Gamma_1, \Gamma_\varphi$). Then the obtained information of a measurement of $\langle \sigma_z^{Q1} \rangle (t)$  is equivalent to a horizontal line in figure \ref{fig 2qubit_energies}\textbf{B}. The positions of the three peaks give the three frequencies in figure \ref{fig 2qubit_energies}\textbf{B} i.e.~the necessary information to obtain the level splitting of the TLS $\delta$ and the qubits-TLS coupling strength $g$ uniquely. Measuring $g$ at several positions above the sample allows the position of the environmental TLS to be obtained from the local minimum of $g$ in figure \ref{fig couplingplot}  i.e.~a single TLS in the substrate can be located.

\begin{figure}
\includegraphics[scale=1]{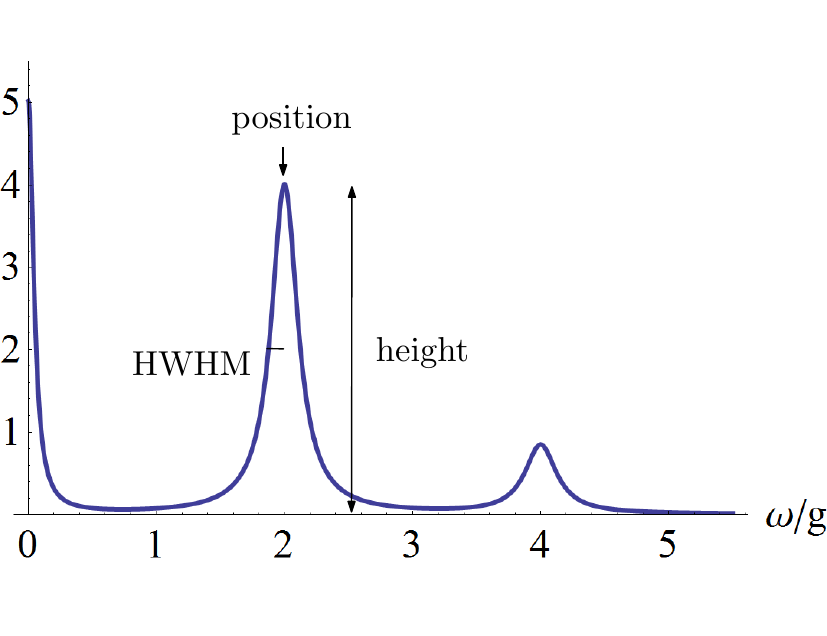} \\%\qquad \qquad
\vspace{0.5cm}
\includegraphics[scale=1]{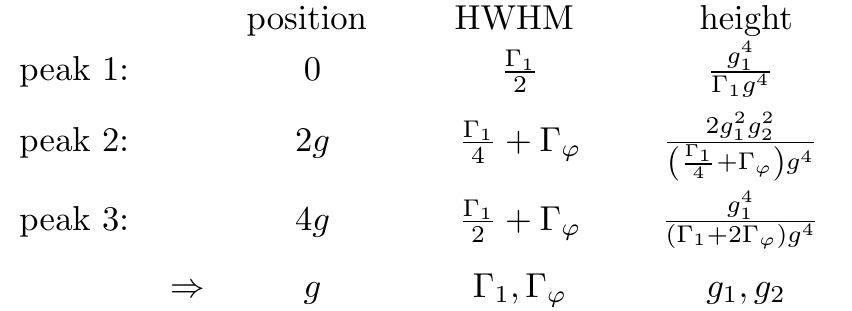}
\caption{Top: Real part of the one-sided Fourier transform of $\langle \sigma_z^{Q1} \rangle (t)$ in the weak decoherence regime (eq. \ref{energies weak decoherence}) for $\delta=0$, Bottom: The table shows how the parameters could be obtained from a measurement of the plot above. \label{fig parameter extraction}}
\end{figure}

The widths and heights of the peaks provide further parameters although they have very complicated dependencies. In the case of resonance ($\delta=0$) however we find three peaks which allow an enormously simplified parameter extraction shown in figure \ref{fig parameter extraction}. Experimentally one such plot provides enough information to obtain all system parameters: $g$ from the position of the peaks, $\Gamma_1$ and $\Gamma_\varphi$ from the HWHM, and (having obtained these three parameters) $g_1$ and $g_2$ from the heights of the peaks. All system parameters can be obtained from one measurement of the time evolution of the excited-state-population of one of the probe qubits on resonance with the TLS. To reach resonance experimentally the qubits could always be tuned to resonance with the TLS, once $\delta$ is obtained as explained in the previous paragraph.

\begin{figure}
\includegraphics[width=0.5 \columnwidth]{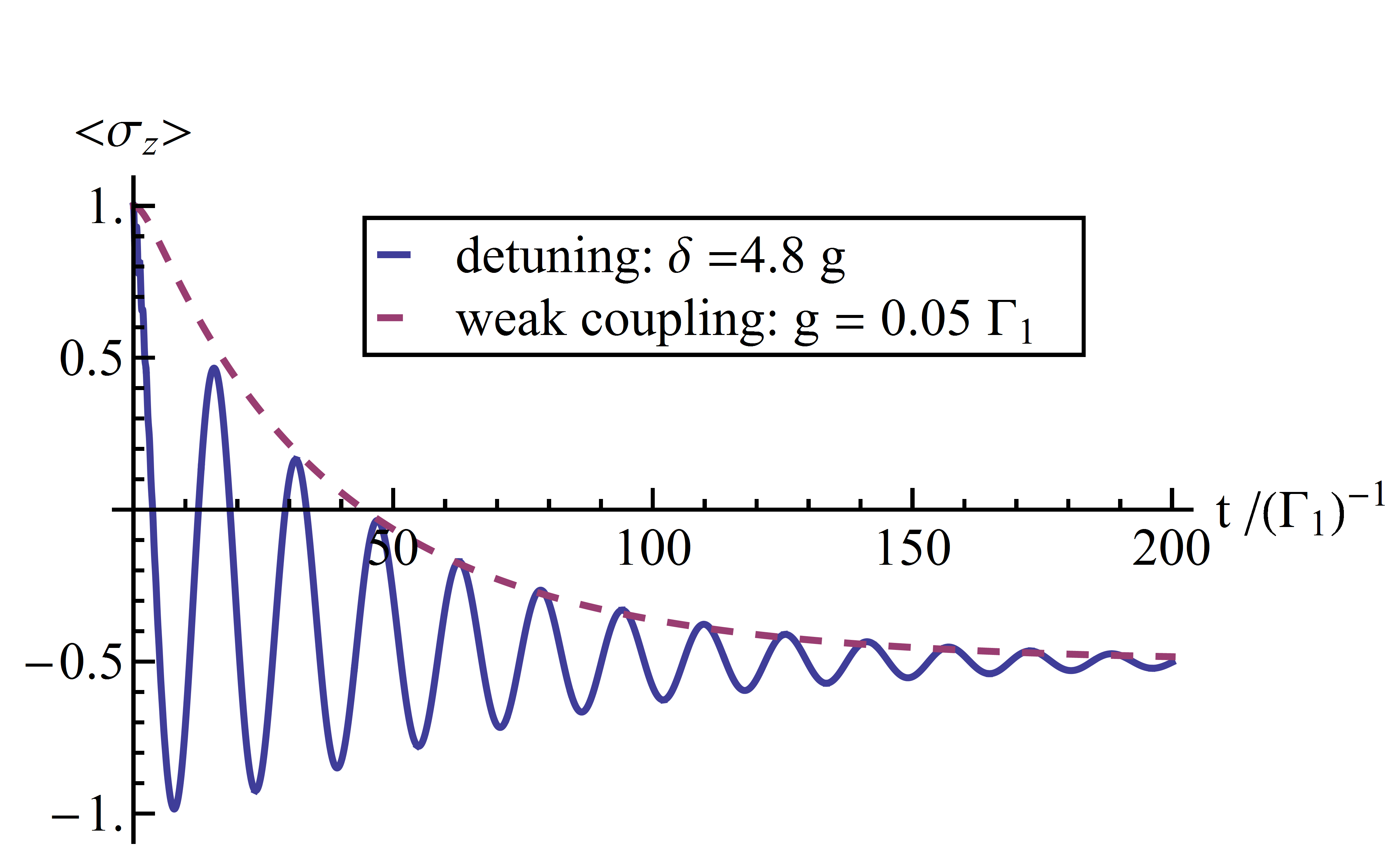}
\includegraphics[width=0.5 \columnwidth]{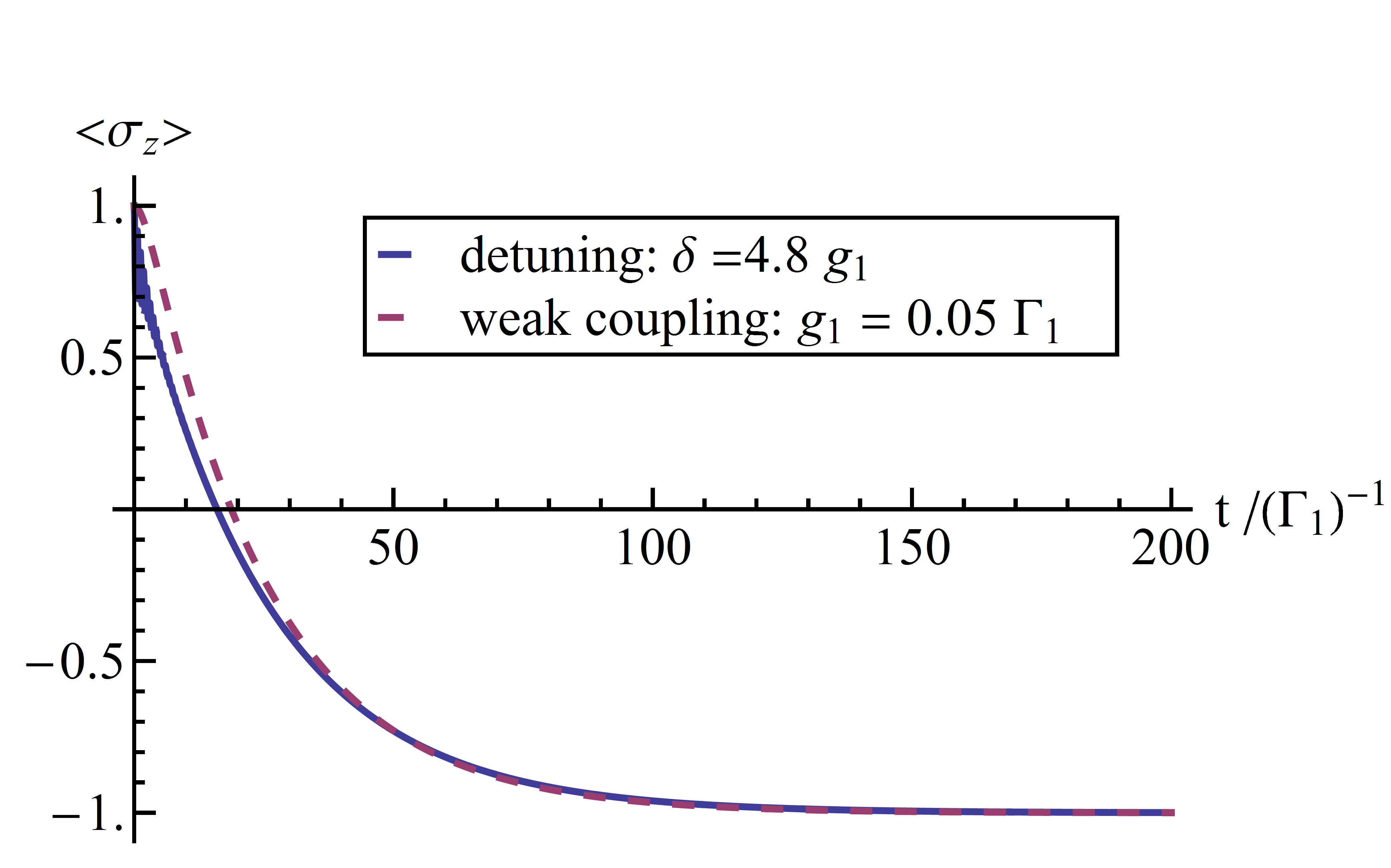}
\caption{Expectation value of qubit 1 for the case of two qubits coupled to a TLS and $g_1=g_2$ (top) and a single qubit coupled to a TLS ($g_2=0$, bottom). Both plots show two different cases: large detuning $\delta=4.8g$ and weak qubit-TLS coupling $g=0.05\Gamma_1$. For all plots $\Gamma_\varphi=0$.  These two different cases can only be clearly distinguished with two qubits. \label{detuning vs weakcoupl}}
\end{figure}

The major difference between the two qubits and a single qubit is the behaviour for detuning to the TLS. While the single qubit is effectively decoupled by detuning, the addition of a second qubit maintains an oscillating signal via the TLS-induced effective coupling between the qubits. As a result strong detuning and weak qubit-TLS coupling show two fundamentally different behaviours and can be distinguished with two qubits (figure \ref{detuning vs weakcoupl}). The system is also sensitive to TLS over a wider frequency range as the TLS-induced coupling decreases more slowly with detuning.

Furthermore the additional two lower frequencies in figure \ref{fig 2qubit_energies}\textbf{B} which correspond to oscillations between the two qubits make it possible to obtain the detuning without changing the level splitting of the qubits.

\subsection{Strong decoherence regime - decaying behaviour}
\begin{figure}
\includegraphics[scale=1]{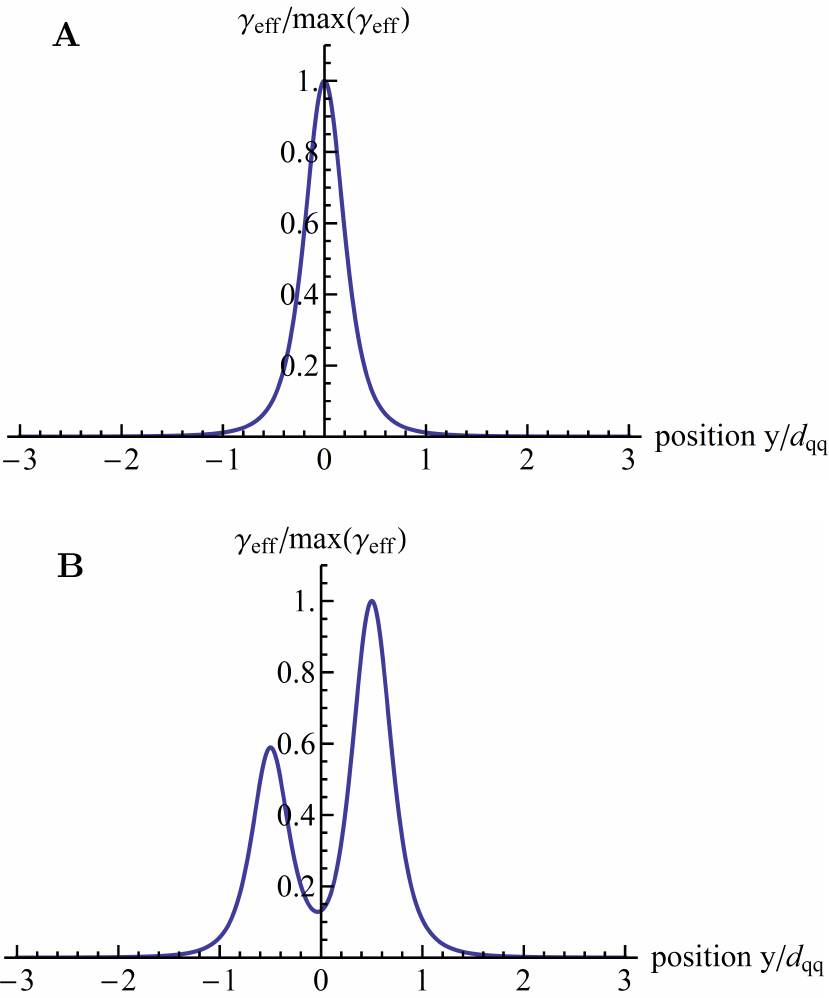}
\caption{\textbf{A} Characteristic behaviour of the effective decay rate of the energy of a single qubit ($g_2=0$) as a function of the position y. This plot is in the decaying regime $\Gamma_1 = 10 g_1$ and $\Gamma_\varphi=0$ .
\textbf{B} Characteristic behaviour of the effective decay rate of the energies of two qubits as a function of the position y. This plot is in the decaying regime $\Gamma_1 = 10 g_1$ and $\Gamma_\varphi=0$ 
\label{fig spatial detection}}
\end{figure}

Scanning a substrate for isolated TLS one might find very different decoherence strengths for each TLS, some of which might be fluctuating so strongly (or coupled so weakly) that no coherent oscillations will occur even when the qubits are directly above it. In that case the above technique of parameter extraction is no longer applicable. However, the TLS can still be located (both with a single and two qubits) by monitoring the decay rate of the qubit at different locations along the y-axis in figure \ref{fig system}\textbf{A}.

For pure relaxation the position dependency of the effective decay rate is shown in figures \ref{fig spatial detection}\textbf{A} (single qubit probe) and \ref{fig spatial detection}\textbf{B} (two qubit probes). The characteristic behaviour provides the position of the TLS.

\section{Experimental realisations \label{experimental realisations}}
Although in principle any qubit architecture can be adapted for performing decoherence microscopy, in order to study microscopic pockets of coherence, atomic scale qubits with long coherence times are ideal.  In solid-state, this implies spin donors or defects, such as semiconductor donors \cite{Kane1998, Sousa2004, Morello2010} or colour centres in diamond \cite{Balasubramanian2008, Balasubramanian2009, Childress2006}.

As the NV centre in diamond is an experimentally established and well investigated system we discuss the requirements to use these centres in a dual-probe configuration.  In recent experiments the intrinsic decoherence of the NV centre is weak and dominated by the dephasing\cite{Balasubramanian2008, Childress2006}, which sets the sensitivity limit for probing environmental pockets of coherence. Isotopic purification of the diamond lattice\cite{Balasubramanian2009} is currently investigated for quantum computation and sensing purposes and will result in much longer coherence times leading to a corresponding increase in sensitivity.

In order to unambiguously identify coupling via the environment, we need to minimise or eliminate coupling between the probes.  This can be achieved in two configurations. Using the nuclear spin states of the Nitrogen within each NV centre as the qubit probes provides a strong electron-nuclear coupling to electron spins in the environment whilst minimising the inter qubit (nuclear-nuclear) coupling.  For both a probe-TLS distance and probe separation of 5 nm, we require a $T_2 >$ 20 ms to reach the probe-TLS oscillation limit. Using current estimates\cite{Balasubramanian2009,Mizuochi2009} for the dephasing channels due to $^{13}$C spins in the diamond lattice, this requires a $^{13}$C concentration below 0.03\%. For this probe separation, the inter-probe coupling is a factor of 1000 times smaller than the probe-TLS coupling and therefore provides no extra complication to the analysis.

A second method of achieving strong probe-TLS coupling is to use a pair of NV centres whose crystallographic orientation is such that the natural NV-NV coupling is eliminated due to the angular dependence of the dipolar interaction.  Although there are more serious fabrication challenges with this configuration, the maximum dephasing required ($T_2 > 20\mu s$) is considerably less due to the strength of the electron-electron interaction. Such dephasing times are well within the range currently seen in experiments using NV centers\cite{Balasubramanian2009}.

In either of these configurations, detecting sample impurities with large magnetic moments such as single molecule magnets ($^8$Fe, Ferritin)\cite{Barco2000, Awschalom1992, Tejada1997} is considerably easier, even with currently available intrinsic decoherence times.  Depending on the background field configuration and qubit operating mode, these impurities will induce either dephasing dominated or energy exchange processes.\\

\section{Conclusion}
In this paper we have investigated the concept of a dual-probe decoherence microscope.  Using a general model, we have studied the key characteristics of such a system analytically and numerically. Mapping out the temporal dynamics of the qubit probes provides detailed information about a TLS, the simplest example of a pocket of coherence contained within the sample. In addition to the TLS' level splitting and the coupling strengths to the probes one can obtain its dephasing and relaxation rate which implies the coupling strength to its surround environment and that environment's constitution. A dual-probe configuration simplifies the measurement process and increases sensitivity to detuned TLS. Furthermore we have shown how the oscillation amplitude of environmentally mediated coupling between two probes is largely unaffected by detuning and decoherence of the mediating TLS, although the frequency of oscillation still depends on detuning. The close relationship between environmentally mediated coupling and non-Markovian dynamics makes a dual-probe configuration ideal for both probing an environment's potential to induce non-Markovian dynamics in a system as well as detecting the spatial extent and interrogate pockets of coherence which sit within a more complex environment.

\ack
We would like to thank N. Oxtoby, S. Huelga and N. Vogt for helpful discussions. This work was supported by the CFN of DFG, the EU project SOLID and the US ARO under contract no.\ W911NF-09-1-0336. We acknowledge support by Deutsche Forschungsgemeinschaft and Open Access Publishing Fund of Karlsruhe Institute of Technology.

\appendix
\section{Analytical understanding of the expectation values and their Fourier transforms\label{app Fourier transform}}
In this appendix we give the three analytical solutions each for no detuning $\delta = 0$ (i.e. $\delta \ll g$). As mentioned before all three solutions were obtained in the subspace of one excitation plus the ground state with the assumption $\omega_q \gg g_1, g_2$. Furthermore a low temperature bath $\omega_q \gg T$ was assumed. First the secular approximation is explained, then the three solutions are given.

\subsection{Matrix form of the master equation and the secular approximation}
The Bloch-Refield equations, eq. \ref{bloch-redfield}, can always be rewritten as a matrix multiplication. To do so, the elements of the density matrix need to be written in a vector. Here we choose the particular order: all diagonal elements first, then all off-diagonal elements. To solve the equations the resulting Redfield tensor $\mathcal{R}$, which is now a matrix, needs to be diagonalized.

The secular approximation means that for this diagonalisation process one can neglect off-diagonal elements in the Redfield tensor $\mathcal{R}$ when there is a large difference between the corresponding diagonal elements. In our chosen order we can regard separate blocks in the Redfield-tensor (see below). The diagonal elements in the lower right block each have a term $-i \omega_{jk}$, whose magnitude is given by the energy difference of the two system states $j$ and $k$. If these level splittings are large compared to the decoherence rates, then the two blocks linking diagonal and off-diagonal density matrix elements can be set to zero, i.e.~the upper right block and the lower left block. If the differences of the energy differences $\omega_{jk} - \omega_{lm}$ are also large compared to the decoherence rates, then all off-diagonal Redfield tensor elements in the lower right block can also be set to zero. This is what we call the full secular approximation. The resulting Redfield tensor is given by:

%\begin{widetext}

\begin{align*}
\left( \begin{array}{c} \dot \rho_{11}\\ \dot \rho_{22} \\ \dot \rho_{33} \\ \vdots \\  \\ \dot \rho_{12} \\ \dot \rho_{13} \\ \vdots \\    \end{array} \right)
= \left(
\begin{array}{cc}
\begin{array}{cccc}
\mathcal{R}_{1111} & \mathcal{R}_{1122} & \mathcal{R}_{1133} & \dots  \\
\mathcal{R}_{2211} & \mathcal{R}_{2222} & \mathcal{R}_{2233} & 		\\
 \mathcal{R}_{3311} & \mathcal{R}_{3322} & \mathcal{R}_{3333} &\\
 \dots &&& \ddots \\
 \end{array}
& \text{{\Huge 0}} \\
\text{{\Huge 0}} &
\begin{array}{ccc}
 \mathcal{R}_{1212} -i \omega_{12} & 0 & 0  \\
 0	& \mathcal{R}_{1313} - i \omega_{13} & 0\\
 0 &0& \ddots
 \end{array}
\end{array}
\right)
\left( \begin{array}{c}  \rho_{11}\\  \rho_{22} \\ \rho_{33} \\ \vdots \\  \\  \rho_{12} \\  \rho_{13} \\ \vdots \\    \end{array} \right)
\end{align*}

Mathematically the secular approximation is analogous to the rotating wave approximation: Separating the Redfield tensor into a coherent part (i.e.~the $-i \omega_{jk}$ terms) and a decoherent part we can define an analogous ``interaction picture'' for the vector of density matrix elements $\vec \rho$:
\begin{align}
&\mathcal{R}=\mathcal{R}^{coh}+\mathcal{R}^{dec}\\
&\vec{\tilde \rho} := \exp (-\mathcal{R}^{coh} t) \vec \rho \\
\Rightarrow \quad &\frac{d}{dt}\vec{\tilde \rho} = \exp (-\mathcal{R}^{coh} t) \mathcal{R}^{dec} \exp (\mathcal{R}^{coh} t) \: \vec{\tilde{\rho}} =: \tilde{\mathcal{R}} \: \vec{\tilde{\rho}}
\end{align}
This results in a new Redfield tensor $\tilde{\mathcal{R}}$ in this ``interaction picture'' where all off-diagonal elements are multiplied by a rotating term with the rotation frequency equal to the difference of the corresponding diagonal $\mathcal{R}$-elements. The secular approximation can then be justified analogously to the rotating wave approximation.

\subsection{Weak decoherence regime $g \gg \Gamma_1, \Gamma_\varphi$ \label{app weak decoherence solutions}}
The two qubits couple to the same environmental TLS strongly compared to the decoherence of the TLS. The full secular approximation is applied.
\begin{align}
\begin{aligned}
\langle \sigma_z^{Q1} \rangle(t) &= \frac{g_2^4-g_1^4-2 g_1^2 g_2^2}{g^4}
 + \frac{g_1^4}{g^4} \;e^{-\frac{\Gamma_1 t}{2}} 
 + \frac{4 g_1^2 g_2^2}{g^4} \;e^{-\frac{\Gamma_1 t}{4}-\Gamma_\varphi t}  \cos\left[2 g\; t\right] 
 + \frac{g_1^4}{g^4}   \;e^{-\frac{\Gamma_1 t}{2}-\Gamma_\varphi t}  \cos\left[4 g\; t\right] 
\\
\langle \sigma_z^{Q2} \rangle(t) &= - \frac{g_1^4+g_2^4}{g^4}
+ \frac{g_1^2 g_2^2}{g^4} e^{-\frac{\Gamma_1 t}{2}} 
- \frac{4 g_1^2 g_2^2}{g^4} e^{\left(-\frac{\Gamma_1}{4}-\Gamma_\varphi\right) t}  \cos\left[2g \; t \right]
+ \frac{g_1^2 g_2^2}{g^4} e^{\left(-\frac{\Gamma_1}{2}-\Gamma_\varphi\right) t} \cos\left[4g \; t\right] 
\\
\langle \sigma_z^{TLS} \rangle(t) &= -1 + \frac{g_1^2}{g^2} e^{-\frac{\Gamma_1 t}{2}} - \frac{g_1^2}{g^2} e^{-\frac{\Gamma_1 t}{2}-\Gamma_\varphi t} \cos\left[4g \;t\right] \end{aligned} \label{energies weak decoherence}
\end{align}

\subsection{Purely transversal TLS-bath coupling $v_\parallel=0 \Rightarrow \Gamma_\varphi=0$\label{app purely transversal coupling solutions}}
The TLS-bath coupling is purely transversal, i.e. there is relaxation only. The secular approximation need not be applied here, due to our particular choice of environmental coupling operator.  Therefore this solution is also valid for strong relaxation.

\begin{multline}
\langle \sigma_z^{Q1} \rangle(t)=\frac{-g_1^4-2 g_1^2 g_2^2+g_2^4}{g^4}
-\frac{64\text{  }g_1^4 }{\mu^2 g^2}e^{-\frac{\Gamma_1 t}{2}}
+\frac{4 g_1^2 g_2^2 }{\mu g^4} e^{-\frac{\Gamma_1 t}{4}}
\left(\mu \cosh\left[\frac{\mu t}{4}\right]+\Gamma_1 \sinh\left[\frac{\mu t}{4}\right]\right)\\
+\frac{2 g_1^4 }{\mu^2 g^4}e^{-\frac{\Gamma_1 t}{2}}
\left(\left(\Gamma_1^2-32 g^2\right) \cosh\left[\frac{\mu t}{2}\right]+\Gamma_1 \mu \sinh\left[\frac{\mu t}{2}\right]\right)
\end{multline}
\begin{multline}
\langle \sigma_z^{Q2} \rangle(t) =-\frac{g_1^4+g_2^4}{g^4}
-\frac{64 g_1^2 g_2^2 }{\mu^2 g^2} e^{-\frac{\Gamma_1 t}{2}}
-\frac{4\text{  }g_1^2 g_2^2 }{\mu g^4}e^{-\frac{\Gamma_1 t}{4}}
\left(\mu \cosh\left[\frac{\mu t}{4}\right]+\Gamma_1 \sinh\left[\frac{\mu t}{4}\right]\right)\\
+\frac{2 g_1^2 g_2^2 }{\mu^2 g^4} e^{-\frac{\Gamma_1 t}{2}}
\left(\left(\Gamma_1^2-32 g^2\right) \cosh\left[\frac{\mu t}{2}\right]+\Gamma_1 \mu \sinh\left[\frac{\mu t}{2}\right]\right)
\end{multline}
\begin{multline}
\langle \sigma_z^{TLS} \rangle(t)=-1+\frac{128 g_1^2}{\Gamma_1^2-64 g^2}e^{-\frac{\Gamma_1 t}{2}} \sinh\left[\frac{1}{4} \mu t\right]^2\\
\end{multline}
where $\mu := \sqrt{\Gamma_1^2-64g^2}$\\

\subsection{Purely longitudinal TLS-bath coupling $v_\perp=0 \Rightarrow \Gamma_1=0$ \label{app purely longitudinal coupling solutions}}
The TLS-bath coupling is purely longitudinal i.e. there is dephasing only. Again the secular approximation need not be applied here, due to our particular choice of initial state and environmental coupling operator.  Therefore this solution is also valid for strong dephasing.

\begin{multline}
\langle \sigma_z^{Q1}\rangle(t) = \frac{-2 g_1^2 g_2^2+g_2^4}{g^4}\\
 +e^{-t \Gamma _{\varphi }}\text{  }\frac{g_1^2}{g^4}\left( 4 g_2^2\cosh\left[\mu _2 t\right]+\frac{4 g_2^2\Gamma _{\varphi }}{\mu _2}\sinh\left[\mu _2 t\right]+g_1^2\cosh\left[\mu _3 t\right] +\frac{g_1^2\Gamma _{\varphi }}{\mu _3}\sinh\left[\mu _3 t\right]\right)
\end{multline}
\begin{multline}
\langle \sigma_z^{Q2}\rangle(t)=-\frac{g_1^4-g_1^2 g_2^2+g_2^4}{g^4}\\
+e^{-t \Gamma _{\varphi }}\text{  }\frac{g_1^2g_2^2}{g^4}\left(-4\cosh\left[\mu _2 t\right]-\frac{4\Gamma _{\varphi }}{\mu _2}\sinh\left[\mu _2 t\right]+\cosh\left[\mu _3 t\right]+\frac{ \Gamma _{\varphi }}{\mu _3}\sinh\left[\mu _3 t\right]\right)
\end{multline}
\begin{multline}
\langle \sigma_z^{TLS}\rangle(t) = - \frac{g_2^2}{g^2} - e^{-t \Gamma _{\varphi }}\text{  }\frac{g_1^2}{g^2}\left(\cosh\left[\mu _3 t\right]+\frac{\Gamma _{\varphi }}{\mu _3}\sinh\left[\mu _3 t\right]\right)\\
\end{multline}
\begin{tabbing}
where  \= $\mu _{2 } := \sqrt{\Gamma _{\varphi }^2-4 g^2}$\\
\> $\mu _3 := \sqrt{\Gamma _{\varphi }^2-16 g^2}$
\end{tabbing}
%\end{widetext}

\subsection{Fourier transform \label{app Fourier transform 4}}
The expressions in all expectation values contain constant and oscillating terms, with associated decay rates. Terms of this form are more easily understood in the Fourier domain.

For measured signals of the form $e^{-a t} \cos [b t]$ the real part of its one-sided Fourier transform yields two Lorentzian peaks at the position of plus and minus the frequency $b$ and with a half width at half maximum (HWHM) which equals the decay rate $a$:
\begin{align}%*}
\Re \left\{ \int_{0}^\infty e^{-i \omega t}  \left[ e^{-a t} \cos(b t) \right] \;dt \right\} =
 \frac{a}{2 \left(a^2+(\omega-b)^2\right)}+\frac{a}{2 \left(a^2+(\omega+b)^2\right)}
\end{align}%*}
Fitting such peaks allows us to experimentally obtain the frequency and decay rate in a precise and simple way, as displayed in figure \ref{fig parameter extraction}. Additionally, the close correspondence to the parameters in the Fourier domain helps to depict frequencies and decay rates at the same time in figures \ref{fig 1qubit_figure}\textbf{C} and \ref{fig 2qubit_energies}\textbf{B}.

\section{Calculation of the effective decay rate of a sum of decaying oscillations \label{app average and envelope}}
In the strong decoherence regime the oscillations are a sum of several exponentials $\sum_j c_j \exp(- \gamma_j t)$. To find \emph{one} effective decay rate we can simply use eq. \ref{effective decay rate calculation}. This procedure is sensible when the different decay rates are not too far (i.e.~not orders of magnitude) apart. Note that eq. \ref{effective decay rate calculation} can also be calculated from an integration:
\begin{align}%*}
\frac{c}{\gamma_{eff}} = \frac{\sum_j c_j}{\gamma_{eff}} = \int_0^\infty \sum_j c_j \exp(- \gamma_j t) \; dt = \sum_j \frac{c_j}{\gamma_j}
\end{align}%*}

In the weak decoherence regime on the other hand we have additional oscillations for several terms, i.e.~an expression of the form 
\begin{align}
\sum_j c_j \exp(- \gamma_j t) cos(\omega_j t)
\label{decaying oscillations}
\end{align}
where some $\omega_j$ might be zero and the cosine function might be replaced by a sine function for some terms. In such a case (as for example displayed in figure \ref{fig 1qubit_figure}\textbf{B} or \ref{fig 2qubit_energies}\textbf{A}) one needs to decide to either take the \emph{average} or the \emph{envelope} of the oscillations. The effective decay rate of the average neglects the oscillating terms completely and can therefore become zero when there are no purely decaying terms in the expression. On the other hand the average is unambiguous while the upper envelope and the lower envelope can lead to different effective decay rates. This is the reason why the average was chosen in figure \ref{fig effectivedecayrate} for the weak decoherence (oscillating) regime.

The calculation of the \emph{average} is performed by neglecting all oscillating terms and calculating eq. \ref{effective decay rate calculation} from the rest. Numerically that is easily done by rewriting all oscillations in eq. \ref{decaying oscillations} as exponentials, which yields an expression of the form:
\begin{align}%*}
\sum_j c_j \exp((-\gamma_j+i \omega_j) t)
\end{align}%*}
Then all terms with a non-zero imaginary part in the exponential rate can be neglected and the effective decay rate can be calculated as:
\begin{align}%*}
\text{average: }&& \gamma_{eff}&= \underbrace{\frac{\sum_k c_k}{\sum_k c_k/ \gamma_k}}_{\text{pure decays}} 
\end{align}%*}
where $k$ sums over all purely decaying terms.

The calculation of the \emph{envelope} is performed by setting all oscillations (including the algebraic sign) to 1 (upper envelope) or -1 (lower envelope). Afterwards eq. \ref{effective decay rate calculation} can be applied to all terms. Numerically that can easily be performed by taking the magnitude of the coefficients and real parts of the rates for all oscillating terms:
\begin{align}%*}
\text{upper envelope: }&& \gamma_{eff}&= \frac{\sum_k c_k+\sum_l |c_l|}{\sum_k c_k/ \gamma_k+\sum_l |c_l|/ \gamma_l}  \\
\text{lower envelope: }&& \gamma_{eff}&= \frac{\sum_k c_k-\sum_l |c_l|}{\sum_k c_k/ \gamma_k-\sum_l |c_l|/ \gamma_l}  
\end{align}%*}
where $k$ sums over all purely decaying terms and $l$ sums over all oscillating terms. For the purely decaying terms it is important not to take the magnitude of the coefficients in case negative coefficients $c_k<0$ occur. 

In principle all terms with a non-zero imaginary part of the exponential rate are oscillations. However, when this imaginary part (which is the angular frequency of the oscillation) is smaller than the real part (which is the decay rate) then this term decays strongly before the time period of one oscillation, i.e. the term looks like a pure (non-exponential) decay. For a numerical criterion whether a term should be categorised as oscillating or purely decaying one could therefore measure the imaginary part relative to the real part for each individual term. However for simplicity of the criterion we categorize all terms with an imaginary part of the exponential rate below 0.1 (where $g=1$) as purely decaying terms in our numerical calculations for figure \ref{fig effectivedecayrate}. This is about one order of magnitude less than the decay rates plotted in figure~\ref{fig effectivedecayrate}.

\section{Hamiltonian eigenstates of the system \label{app eigenstates}}
For our system of two qubits coupled to one TLS the unnormalised eigenstates indicated in figure \ref{fig system}\textbf{C} are: %file: B-R 2q1t subspace-reduction justification.nb
\begin{align}
\begin{aligned}
|8\rangle &=|\uparrow \uparrow \uparrow \rangle\\
|7\rangle &=(-\delta + \sqrt{\delta^2+4g^2}) |\uparrow \uparrow \downarrow \rangle + 2 g_2 | \uparrow \downarrow \uparrow \rangle + 2 g_1 | \downarrow \uparrow \uparrow \rangle\\
|6\rangle& =-g_1 |\uparrow \downarrow \uparrow \rangle + g_2 | \downarrow \uparrow \uparrow \rangle\\
|5\rangle& =(-\delta - \sqrt{\delta^2+4g^2}) | \uparrow \uparrow \downarrow \rangle + 2 g_2 | \uparrow \downarrow \uparrow \rangle + 2 g_1 | \downarrow \uparrow \uparrow \rangle\\
|4\rangle& =2 g_1 |\uparrow \downarrow \downarrow \rangle + 2 g_2 | \downarrow \uparrow \downarrow \rangle + (\delta +\sqrt{\delta^2+4g^2}) |\downarrow \downarrow\uparrow \rangle\\
|3\rangle& =-g_2 |\uparrow \downarrow \downarrow \rangle + g_1 | \downarrow \uparrow \downarrow \rangle\\
|2\rangle& =2 g_1 | \uparrow \downarrow \downarrow \rangle + 2 g_2 | \downarrow \uparrow \downarrow \rangle + (\delta - \sqrt{\delta^2+4g^2}) | \downarrow \downarrow \uparrow \rangle\\
|1\rangle& =|\downarrow \downarrow \downarrow \rangle
\end{aligned}
\end{align}
where $\uparrow$ indicates an excited state and $\downarrow$ a ground state of the two qubits and the TLS in the order $|Q1,Q2,TLS\rangle$.\\

Setting $g_2 = 0$ and tracing out the second qubit yields the system of only one qubit coupled to a TLS discussed in section \ref{one qubit}. Performing these operations on the above states we find:
\begin{align}
\begin{aligned}
|8\rangle &\rightarrow |8\rangle_{1q} = |\uparrow \uparrow \rangle\\
|7\rangle &\rightarrow |7\rangle_{1q} = (-\delta + \sqrt{\delta^2+4g^2}) |\uparrow  \downarrow \rangle + 2 g_1 | \downarrow  \uparrow \rangle\\
|6\rangle &\rightarrow |6\rangle_{1q} = -g_1 |\uparrow \uparrow \rangle \\
|5\rangle &\rightarrow |5\rangle_{1q} = (-\delta - \sqrt{\delta^2+4g^2}) | \uparrow  \downarrow \rangle  + 2 g_1 | \downarrow  \uparrow \rangle\\
|4\rangle &\rightarrow |4\rangle_{1q} = 2 g_1 |\uparrow  \downarrow \rangle + (\delta +\sqrt{\delta^2+4g^2}) |\downarrow \uparrow \rangle\\
|3\rangle &\rightarrow |3\rangle_{1q} = g_1 | \downarrow  \downarrow \rangle\\
|2\rangle &\rightarrow |2\rangle_{1q} = 2 g_1 | \uparrow  \downarrow \rangle + (\delta - \sqrt{\delta^2+4g^2}) | \downarrow  \uparrow \rangle\\
|1\rangle &\rightarrow |1\rangle_{1q} = |\downarrow \downarrow \rangle
\end{aligned}
\end{align}
Several states become equivalent:
\begin{align}
\ket{3}_{1q}&=g_1 \ket{1}_{1q}\\
\ket{4}_{1q}&=\frac{2 g_1 }{-\delta + \sqrt{\delta^2+4g^2}}\ket{7}_{1q}\\
\ket{5}_{1q}&=\frac{-\delta-\sqrt{\delta^2+4 g_1^2}}{2 g_1} \ket{2}_{1q}\\
\ket{6}_{1q}&=-g_1 \ket{8}_{1q}
\end{align}
We therefore need (as one should expect) only four states to describe this reduced system.

\section*{References}
\bibliographystyle{unsrt}
\bibliography{publication}

\begin{thebibliography}{10}

\bibitem{ShnirmanSchoen2008}
Alexander Shnirman, Gerd Sch\"on, Ivar Martin, and Yuriy Makhlin.
\newblock Josephson qubits as probes of 1/f noise.
\newblock In {\em "Electron Correlation in New Materials and Nanosystems"},
  number 241 in NATO Science Series, pages 343--356. NATO, Springer, 2007.

\bibitem{Coledecoherencemicroscopy}
Jared~H. Cole and Lloyd C.~L. Hollenberg.
\newblock Scanning quantum decoherence microscopy.
\newblock {\em Nanotechnology}, 20(49):495401, DEC 9 2009.

\bibitem{Chernobrod2005}
Boris~M. Chernobrod and Gennady~P. Berman.
\newblock Spin microscope based on optically detected magnetic resonance.
\newblock {\em Journal of Applied Physics}, 97(1):014903, 2005.

\bibitem{Weber2008}
Jochen Weber, Juergen Weis, Maik Hauser, and Klaus Von~Klitzing.
\newblock {Fabrication of an array of single-electron transistors for a
  scanning probe microscope sensor}.
\newblock {\em Nanotechnology}, 19(37):375301, {SEP 17} 2008.

\bibitem{Sousa2009}
Rogerio de~Sousa.
\newblock {Electron Spin as a Spectrometer of Nuclear-Spin Noise and Other
  Fluctuations}.
\newblock In {Fanciulli, M}, editor, {\em {Electron spin resonance and related
  phenomena in low-dimensional structures}}, volume {115} of {\em {Topics in
  Applied Physics}}, pages {183--220}, {2009}.
\newblock {International Workshop on Electron Spin Resonance and Related
  Phenomena in Low-Dimensional Structures, Sanremo, Italy, March 06-08, 2006}.

\bibitem{Martinis2005}
John~M. Martinis, K.~B. Cooper, R.~McDermott, Matthias Steffen, Markus Ansmann,
  K.~D. Osborn, K.~Cicak, Seongshik Oh, D.~P. Pappas, R.~W. Simmonds, and
  Clare~C. Yu.
\newblock Decoherence in josephson qubits from dielectric loss.
\newblock {\em Physical Review Letters}, 95:210503, Nov 2005.

\bibitem{Neeley2008}
Matthew Neeley, M.~Ansmann, Radoslaw~C. Bialczak, M.~Hofheinz, N.~Katz, Erik
  Lucero, A.~O'Connell, H.~Wang, A.~N. Cleland, and John~M. Martinis.
\newblock Process tomography of quantum memory in a josephson-phase qubit
  coupledto a two-level state.
\newblock {\em Nature Physics}, 4(7):523--526, {JUL} 2008.

\bibitem{Martin2005}
I.~Martin, L.~Bulaevskii, and A.~Shnirman.
\newblock Tunneling spectroscopy of two-level systems inside a josephson
  junction.
\newblock {\em Physical Review Letters}, 95:127002, Sep 2005.

\bibitem{Bushev2010}
P.~Bushev, C.~Mueller, J.~Lisenfeld, J.~H. Cole, A.~Lukashenko, A.~Shnirman,
  and A.~V. Ustinov.
\newblock Multiphoton spectroscopy of a hybrid quantum system.
\newblock {\em Physical Review B}, 82(13):134530, OCT 25 2010.

\bibitem{Cole2010defectmodels}
J.~H. Cole, C.~M\"{u}ller, P.~Bushev, G.~J. Grabovskij, J.~Lisenfeld,
  A.~Lukashenko, A.~V. Ustinov, and A.~Shnirman.
\newblock Quantitative evaluation of defect-models in superconducting phase
  qubits.
\newblock {\em Applied Physics Letters}, 97(25):252501, 2010.

\bibitem{Lisenfeld2010}
J\"urgen Lisenfeld, Clemens M\"uller, Jared~H. Cole, Pavel Bushev, Alexander
  Lukashenko, Alexander Shnirman, and Alexey~V. Ustinov.
\newblock Rabi spectroscopy of a qubit-fluctuator system.
\newblock {\em Physical Review B}, 81(10):100511, Mar 2010.

\bibitem{Jelezko2004}
F.~Jelezko, T.~Gaebel, I.~Popa, A.~Gruber, and J.~Wrachtrup.
\newblock Observation of coherent oscillations in a single electron spin.
\newblock {\em Physical Review Letters}, 92:076401, Feb 2004.

\bibitem{Jelezko2004a}
F.~Jelezko, T.~Gaebel, I.~Popa, M.~Domhan, A.~Gruber, and J.~Wrachtrup.
\newblock Observation of coherent oscillation of a single nuclear spin and
  realization of a two-qubit conditional quantum gate.
\newblock {\em Physical Review Letters}, 93:130501, Sep 2004.

\bibitem{Gaebel2006}
T~Gaebel, M~Domhan, I~Popa, C~Wittmann, P~Neumann, F~Jelezko, JR~Rabeau,
  N~Stavrias, AD~Greentree, S~Prawer, J~Meijer, J~Twamley, PR~Hemmer, and
  J~Wrachtrup.
\newblock Room-temperature coherent coupling of single spins in diamond.
\newblock {\em Nature Physics}, 2(6):408--413, {JUN} 2006.

\bibitem{Childress2006}
L.~Childress, M.~V.~Gurudev Dutt, J.~M. Taylor, A.~S. Zibrov, F.~Jelezko,
  J.~Wrachtrup, P.~R. Hemmer, and M.~D. Lukin.
\newblock Coherent dynamics of coupled electron and nuclear spin qubits in
  diamond.
\newblock {\em Science}, 314(5797):281--285, {OCT 13} 2006.

\bibitem{Morello2010}
Andrea Morello, Jarryd~J. Pla, Floris~A. Zwanenburg, Kok~W. Chan, Kuan~Y. Tan,
  Hans Huebl, Mikko Mottonen, Christopher~D. Nugroho, Changyi Yang, Jessica~A.
  van Donkelaar, Andrew D.~C. Alves, David~N. Jamieson, Christopher~C. Escott,
  Lloyd C.~L. Hollenberg, Robert~G. Clark, and Andrew~S. Dzurak.
\newblock Single-shot readout of an electron spin in silicon.
\newblock {\em Nature}, 467(7316):687--691, OCT 7 2010.

\bibitem{Barco2000}
E.~del Barco, J.~M. Hernandez, J.~Tejada, N.~Biskup, R.~Achey, I.~Rutel,
  N.~Dalal, and J.~Brooks.
\newblock High-frequency resonant experiments in $fe_{8}$ molecular clusters.
\newblock {\em Physical Review B}, 62(5):3018--3021, Aug 2000.

\bibitem{Awschalom1992}
D.~D. Awschalom, J.~F. Smyth, G.~Grinstein, D.~P. DiVincenzo, and D.~Loss.
\newblock Macroscopic quantum tunneling in magnetic proteins.
\newblock {\em Physical Review Letters}, 68(20):3092--3095, May 1992.

\bibitem{Tejada1997}
J.~Tejada, X.~X. Zhang, E.~del Barco, J.~M. Hern\'andez, and E.~M. Chudnovsky.
\newblock Macroscopic resonant tunneling of magnetization in ferritin.
\newblock {\em Physical Review Letters}, 79(9):1754--1757, Sep 1997.

\bibitem{Balasubramanian2008}
Gopalakrishnan Balasubramanian, I.~Y. Chan, Roman Kolesov, Mohannad Al-Hmoud,
  Julia Tisler, Chang Shin, Changdong Kim, Aleksander Wojcik, Philip~R. Hemmer,
  Anke Krueger, Tobias Hanke, Alfred Leitenstorfer, Rudolf Bratschitsch, Fedor
  Jelezko, and Joerg Wrachtrup.
\newblock Nanoscale imaging magnetometry with diamond spins under ambient
  conditions.
\newblock {\em Nature}, 455({7213}):648--U46, {OCT 2} 2008.

\bibitem{Maze2008}
J.~R. Maze, P.~L. Stanwix, J.~S. Hodges, S.~Hong, J.~M. Taylor, P.~Cappellaro,
  L.~Jiang, M.~V.~Gurudev Dutt, E.~Togan, A.~S. Zibrov, A.~Yacoby, R.~L.
  Walsworth, and M.~D. Lukin.
\newblock {Nanoscale magnetic sensing with an individual electronic spin in
  diamond}.
\newblock {\em Nature}, 455(7213):644--647, October 2008.

\bibitem{Degen2008}
C.~L. Degen.
\newblock Scanning magnetic field microscope with a diamond single-spin sensor.
\newblock {\em Applied Physics Letters}, 92(24):243111, 2008.

\bibitem{Taylor2008}
J.~M. Taylor, P.~Cappellaro, L.~Childress, L.~Jiang, D.~Budker, P.~R. Hemmer,
  A.~Yacoby, R.~Walsworth, and M.~D. Lukin.
\newblock High-sensitivity diamond magnetometer with nanoscale resolution.
\newblock {\em Nature Physics}, {4}({10}):810--816, {OCT} {2008}.

\bibitem{Balasubramanian2009}
Gopalakrishnan Balasubramanian, Philipp Neumann, Daniel Twitchen, Matthew
  Markham, Roman Kolesov, Norikazu Mizuochi, Junichi Isoya, Jocelyn Achard,
  Johannes Beck, Julia Tissler, Vincent Jacques, Philip~R. Hemmer, Fedor
  Jelezko, and Joerg Wrachtrup.
\newblock Ultralong spin coherence time in isotopically engineered diamond.
\newblock {\em Nature Materials}, 8(5):383--387, {MAY} 2009.

\bibitem{Hall2009}
L.~T. Hall, J.~H. Cole, C.~D. Hill, and L.~C.~L. Hollenberg.
\newblock Sensing of fluctuating nanoscale magnetic fields using
  nitrogen-vacancy centers in diamond.
\newblock {\em Physical Review Letters}, 103:220802, Nov 2009.

\bibitem{Hall2010}
Liam~T. Hall, Charles~D. Hill, Jared~H. Cole, Brigitte Staedler, Frank Caruso,
  Paul Mulvaney, Joerg Wrachtrup, and Lloyd C.~L. Hollenberg.
\newblock Monitoring ion-channel function in real time through quantum
  decoherence.
\newblock {\em Proceedings of the National Academy of Sciences of the United
  States of America}, 107(44):18777--18782, {NOV 2} 2010.

\bibitem{Mueller2009}
Clemens Mueller, Alexander Shnirman, and Yuriy Makhlin.
\newblock {Relaxation of Josephson qubits due to strong coupling to two-level
  systems}.
\newblock {\em Physical Review B}, {80}({13}):94, {OCT} {2009}.

\bibitem{Oxtoby2009}
Neil~P. Oxtoby, Angel Rivas, Susana~F. Huelga, and Rosario Fazio.
\newblock Probing a composite spin-boson environment.
\newblock {\em New Journal of Physics}, 11(6):063028, 2009.

\bibitem{Yuan2008}
Shengjun Yuan, Mikhail~I. Katsnelson, and Hans De~Raedt.
\newblock Decoherence by a spin thermal bath: Role of spin-spin interactions
  and initial state of the bath.
\newblock {\em Physical Review B}, 77(18):184301, May 2008.

\bibitem{Emary2008}
Clive Emary.
\newblock Quantum dynamics in nonequilibrium environments.
\newblock {\em Physical Review A}, 78(3):032105, Sep 2008.

\bibitem{Paladino2008}
E.~Paladino, M.~Sassetti, G.~Falci, and U.~Weiss.
\newblock Characterization of coherent impurity effects in solid-state qubits.
\newblock {\em Physical Review B}, 77(4):041303, Jan 2008.

\bibitem{Raimond2001}
J.~M. Raimond, M.~Brune, and S.~Haroche.
\newblock Manipulating quantum entanglement with atoms and photons in a cavity.
\newblock {\em Review of Modern Physics}, 73(3):565--582, Aug 2001.

\bibitem{Sorensen1999}
Anders S\o{}rensen and Klaus M\o{}lmer.
\newblock Quantum computation with ions in thermal motion.
\newblock {\em Physical Review Letters}, 82(9):1971--1974, Mar 1999.

\bibitem{Zheng2000}
Shi-Biao Zheng and Guang-Can Guo.
\newblock Efficient scheme for two-atom entanglement and quantum information
  processing in cavity qed.
\newblock {\em Physical Review Letters}, 85(11):2392--2395, Sep 2000.

\bibitem{Majer2007}
J.~Majer, J.~M. Chow, J.~M. Gambetta, Jens Koch, B.~R. Johnson, J.~A. Schreier,
  L.~Frunzio, D.~I. Schuster, A.~A. Houck, A.~Wallraff, A.~Blais, M.~H.
  Devoret, S.~M. Girvin, and R.~J. Schoelkopf.
\newblock Coupling superconducting qubits via a cavity bus.
\newblock {\em Nature}, 449(7161):443--447, September 2007.

\bibitem{Filipp2009}
S.~Filipp, P.~Maurer, P.~J. Leek, M.~Baur, R.~Bianchetti, J.~M. Fink,
  M.~G\"oppl, L.~Steffen, J.~M. Gambetta, A.~Blais, and A.~Wallraff.
\newblock Two-qubit state tomography using a joint dispersive readout.
\newblock {\em Phys. Rev. Lett.}, 102:200402, May 2009.

\bibitem{Niskanen2006}
Antti~O. Niskanen, Yasunobu Nakamura, and Jaw-Shen Tsai.
\newblock Tunable coupling scheme for flux qubits at the optimal point.
\newblock {\em Phys. Rev. B}, 73:094506, Mar 2006.

\bibitem{Niemczyk2010}
T.~Niemczyk, F.~Deppe, H.~Huebl, E.~P. Menzel, F.~Hocke, M.~J. Schwarz, J.~J.
  Garcia-Ripoll, D.~Zueco, T.~Hummer, E.~Solano, A.~Marx, and R.~Gross.
\newblock Circuit quantum electrodynamics in the ultrastrong-coupling regime.
\newblock {\em Nat Phys}, 6(10):772--776, October 2010.

\bibitem{Fink2009}
J.~M. Fink, R.~Bianchetti, M.~Baur, M.~G\"oppl, L.~Steffen, S.~Filipp, P.~J.
  Leek, A.~Blais, and A.~Wallraff.
\newblock Dressed collective qubit states and the tavis-cummings model in
  circuit qed.
\newblock {\em Phys. Rev. Lett.}, 103:083601, Aug 2009.

\bibitem{Casanova2010}
J.~Casanova, G.~Romero, I.~Lizuain, J.~J. Garc\'ia-Ripoll, and E.~Solano.
\newblock Deep strong coupling regime of the jaynes-cummings model.
\newblock {\em Phys. Rev. Lett.}, 105:263603, Dec 2010.

\bibitem{Yu2009}
Ting Yu and J.~H. Eberly.
\newblock Sudden death of entanglement.
\newblock {\em Science}, 323(5914):598--601, 2009.

\bibitem{Yonac2006}
Muhammed Yonac, Ting Yu, and J~H Eberly.
\newblock Sudden death of entanglement of two jaynesâ€“cummings atoms.
\newblock {\em Journal of Physics B: Atomic, Molecular and Optical Physics},
  39(15):S621, 2006.

\bibitem{Yu2007}
Ting Yu and J.~H. Eberly.
\newblock Negative entanglement measure, and what it implies.
\newblock {\em J.MOD.OPT.}, 54:2289, 2007.

\bibitem{Cole2010entanglementsuddendeath}
Jared~H Cole.
\newblock Understanding entanglement sudden death through multipartite
  entanglement and quantum correlations.
\newblock {\em Journal of Physics A: Mathematical and Theoretical},
  43(13):135301, 2010.

\bibitem{Ashhab2010}
S.~Ashhab and Franco Nori.
\newblock Qubit-oscillator systems in the ultrastrong-coupling regime and their
  potential for preparing nonclassical states.
\newblock {\em Phys. Rev. A}, 81:042311, Apr 2010.

\bibitem{Bloch1957}
F.~Bloch.
\newblock Generalized theory of relaxation.
\newblock {\em Physical Review}, 105(4):1206--1222, Feb 1957.

\bibitem{Redfield}
A.~G. Redfield.
\newblock On the theory of relaxation processes.
\newblock {\em IBM Journal of Research and Development}, 1(1):19 --31, 1957.

\bibitem{Boissoneault2009}
Maxime Boissonneault, J.~M. Gambetta, and Alexandre Blais.
\newblock Dispersive regime of circuit qed: Photon-dependent qubit dephasing
  and relaxation rates.
\newblock {\em Phys. Rev. A}, 79:013819, Jan 2009.

\bibitem{Breuerbook}
Heinz-Peter Breuer and Francesco Petruccione.
\newblock {\em The theory of open quantum systems}.
\newblock Oxford University Press, 2003.

\bibitem{Lindblad1976}
G.~Lindblad.
\newblock On the generators of quantum dynamical semigroups.
\newblock {\em Communications in Mathematical Physics}, 48:119--130, 1976.
\newblock 10.1007/BF01608499.

\bibitem{Scully1999}
Marlan~O. Scully and M.~Suhail Zubairy.
\newblock {\em Quantum optics}.
\newblock Cambridge Univ. Press, 1999.

\bibitem{gerrybook}
C.C. Gerry and P.L. Knight.
\newblock {\em Introductory quantum optics}.
\newblock Cambridge University Press, 2005.

\bibitem{Crubellier1985}
A~Crubellier, S~Liberman, D~Pavolini, and P~Pillet.
\newblock Superradiance and subradiance. i. interatomic interference and
  symmetry properties in three-level systems.
\newblock {\em Journal of Physics B}, 18(18):3811, 1985.

\bibitem{Wootters98}
William~K. Wootters.
\newblock Entanglement of formation of an arbitrary state of two qubits.
\newblock {\em Physical Review Letters}, 80:2245--2248, Mar 1998.

\bibitem{Cui2008}
Wei Cui, Zai~Rong Xi, and Yu~Pan.
\newblock Optimal decoherence control in non-markovian open dissipative quantum
  systems.
\newblock {\em Physical Review A}, 77:032117, Mar 2008.

\bibitem{Breuer2009}
Heinz-Peter Breuer, Elsi-Mari Laine, and Jyrki Piilo.
\newblock Measure for the degree of non-markovian behavior of quantum processes
  in open systems.
\newblock {\em Physical Review Letters}, 103(21):210401, Nov 2009.

\bibitem{Rivas2010}
\'Angel Rivas, Susana~F. Huelga, and Martin~B. Plenio.
\newblock Entanglement and non-markovianity of quantum evolutions.
\newblock {\em Physical Review Letters}, 105(5):050403, Jul 2010.

\bibitem{Lu2010}
Xiao-Ming Lu, Xiaoguang Wang, and C.~P. Sun.
\newblock Quantum fisher information flow and non-markovian processes of open
  systems.
\newblock {\em Physical Review A}, 82(4):042103, Oct 2010.

\bibitem{Kane1998}
BE~Kane.
\newblock A silicon-based nuclear spin quantum computer.
\newblock {\em Nature}, 393(6681):133--137, MAY 14 1998.

\bibitem{Sousa2004}
Rogerio de~Sousa, J.~D. Delgado, and S.~Das~Sarma.
\newblock Silicon quantum computation based on magnetic dipolar coupling.
\newblock {\em Physical Review A}, 70(5):052304, Nov 2004.

\bibitem{Mizuochi2009}
N.~Mizuochi, P.~Neumann, F.~Rempp, J.~Beck, V.~Jacques, P.~Siyushev,
  K.~Nakamura, D.~J. Twitchen, H.~Watanabe, S.~Yamasaki, F.~Jelezko, and
  J.~Wrachtrup.
\newblock Coherence of single spins coupled to a nuclear spin bath of varying
  density.
\newblock {\em Physical Review B}, 80(4):041201, Jul 2009.

\end{thebibliography}

\end{document}